\documentclass[aps,nofootinbib,prd]{revtex4} 
\usepackage{amsfonts}
\usepackage{amssymb}
\usepackage{graphicx}
\usepackage{pstricks}
\usepackage{epsfig}

\newcommand{\be}{\begin{equation}}
\newcommand{\ee}{\end{equation}}
\newcommand{\beq}{\begin{eqnarray}}
\newcommand{\eeq}{\end{eqnarray}}

\newcommand{\Ree }{{\rm Re\ }}
\newcommand{\Imm }{{\rm Im\ }}

\newcommand{\md}{{\mathrm d}}

\begin{document}

\title{The high-density regime of kinetic-dominated loop quantum cosmology}

\author{M.~Bojowald\footnote{bojowald@gravity.psu.edu}}
\author{W.~Nelson\footnote{nelson@gravity.psu.edu}}

\affiliation{Institute for Gravitation and the Cosmos, Penn State University, State College, PA 16801, U.S.A. } 

\author{D.~Mulryne\footnote{d.mulryne@imperial.ac.uk}}
\affiliation{Department of Physics, Imperial College London, South Kensington campus, London, SW7 2AZ. U.K. }

\author{R.~Tavakol\footnote{r.tavakol@qmul.ac.uk}}
\affiliation{School of Mathematical Sciences, Queen Mary University of London, London E1 4NS, U.K.}

\begin{abstract}
We study the dynamics of states perturbatively expanded about a
harmonic system of loop quantum cosmology, exhibiting a bounce. In
particular, the evolution equations for the first and second order
moments of the system are analyzed. These moments back-react on the
trajectories of the expectation values of the state and hence alter
the energy density at the bounce. This analysis is performed for
isotropic loop quantum cosmology coupled to a scalar field with a
small but non-zero constant potential, hence in a regime in which the
kinetic energy of matter dominates.  Analytic restrictions on the
existence of dynamical coherent states and the meaning of
semi-classicality within these systems are discussed.  A numerical
investigation of the trajectories of states that remain semi-classical
across the bounce demonstrates that, at least for such states, the
bounce persists and that its properties are similar to the standard
case, in which the moments of the states are entirely
neglected. However the bounce density does change, implying that a
quantum bounce may not be guaranteed to happen when the potential is
no longer negligible.
\end{abstract}

\maketitle
%
%
%%%%%%%%%%%%%%%%%%%%%%%%%%%%%%%%%%%%%%%%%%%%%%%%%%%%%%%%%%%%%%%%%%%%%%%%%%%%%%%%%%%%%%%%%%%%%%%%%%%%%%%
%
%
\section{Introduction}\label{sec:intro}

Loop quantum cosmology \cite{LivRev} imports mathematical
constructions and results obtained in the general setting of quantum
geometry realized in loop quantum gravity \cite{Rov,ThomasRev,ALRev}
into quantum cosmology. As a result, the quantum representation is
unitarily inequivalent to the one used in Wheeler--DeWitt
quantizations of cosmological models, a formal difference entailing
also dynamical changes in the behavior of models for the very early
universe. Loop quantum cosmology thus allows one to explore the
physical implications of quantum geometry and its unconventional
space-time structures in a tractable setting.

One of the main changes implied by loop quantum gravity is the use of
holonomies \cite{LoopRep}, objects of the form $\exp(i {\rm curv})$
replacing the usual curvature expressions ``${\rm curv}$'' (extrinsic
and intrinsic curvature, or space-time curvature components) in the
classical equations. The form of the curvature functional
``${\rm curv}$'' appearing in holonomies depends on the specific
quantization of the Hamiltonian constraint used \cite{QSDI} as well as
on details of the reduction to a symmetric context within quantum
gravity \cite{SymmRed}; however, within isotropic cosmological models
it is a function only of the scale factor and its time
derivative. Moreover, neither is the full theory of loop quantum
gravity uniquely and unambiguously formulated yet, nor is the
reduction from general physical states to isotropic ones fully
understood at a quantitative level. Despite such ambiguities, several
general conclusions can be inferred from mathematical properties of
the curvature--holonomy replacement, most importantly related to the
boundedness of holonomy expressions in contrast to the curvature
components. For instance, the Friedmann equation relates extrinsic
curvature $a\dot{a}$ of spatial slices in a dynamical isotropic
universe to the energy density of matter. If extrinsic curvature in
the equation is replaced by a bounded function of a certain form,
boundedness of the energy density automatically follows in spite of
its classical divergence at the big bang singularity.

Replacing curvature expressions by bounded functions is not simply
done by hand but is motivated by quantum theory. Quantization also
means that one is dealing with states rather than classical geometries
subject to the usual evolution equations. Fundamentally, the use of
holonomies primarily implies that the dynamics of quantum states of
universe models is governed not by the differential Wheeler--DeWitt
equation \cite{DeWitt,QCReview}, but by a difference equation
\cite{cosmoIV,IsoCosmo}. These discrete dynamics turn out to be
non-singular \cite{Sing,BSCG}: wave functions extend uniquely across
the classically singular big bang. But while the use of holonomies and
the discreteness of the difference equation implied by it are
important for this result, detailed properties of the transition
depend on various kinds of quantum effects that the use of the wave
function introduces. In particular, in addition to quantum geometry
effects such as discreteness, there are in general quantum
back-reaction effects that characterize deviations of the quantum
evolution of expectation values from the classical evolution caused by
a changing wave function. Simple intuitive boundedness results on
energy densities can be established only if quantum geometry effects
can be shown not to be overpowered by potentially adverse quantum
back-reaction effects. In loop quantum cosmology, this competition of
different quantum effects has not been explored sufficiently in
generic regimes. (Examples in Wheeler--DeWitt quantum cosmology have
shown the sensitivity of some quantum effects related to singularity
avoidance to state properties
\cite{GaussianBohmQC,NonSingBohmQC,WDWBounce}.)

Typically attention has been focused on harmonic models, in which the
quantum back-reaction is entirely absent. This has the advantage of
being analytically tractable and has been shown to generically lead to
bouncing solutions. The robustness of these results for models that
deviate from these special cases has only begun to be explored and in
this paper we further develop a systematic perturbation theory
approach to this question~\cite{EffAc,Karpacz}. Key results reported
here include the interplay of reality conditions (physical
normalization of states) and uncertainty relations, which in
non-harmonic cases turns out to be important for consistent evolution,
numerical or otherwise. By numerical studies, we demonstrate the
sensitivity of bounce properties such as the density to quantum
effects. Also the high sensitivity of evolution through the
bounce on initial values, observed for harmonic models based on an
analysis of dynamical coherent states \cite{BeforeBB,Harmonic}, is
qualitatively confirmed in our more general setting. Although the
leading-order perturbations used here do not allow quantum dynamical
corrections large enough to (potentially) remove the bounce induced by
the holonomy modification, the sensitivity we find appears
sufficiently strong to warrant caution about extrapolations of the
kinetic-dominated behavior to models with significant matter
potentials.
%
%
%%%%%%%%%%%%%%%%%%%%%%%%%%%%%%%%%%%%%%%%%%%%%%%%%%%%%%%%%%%%%%%%%%%%%%%%%%%%%%%%%%%%%%%%%%%%%%%%%%%%%%%
%
%
\section{Loop quantum cosmology}

According to the holonomy replacement scheme, one of the basic objects
in loop quantum cosmology is the holonomy $\exp(i{\rm curv})$ with an
expression ``${\rm curv}$'' linear in curvature components. For an
isotropic cosmological model, this means ${\rm curv}= g(a)\dot{a}$ in
terms of the scale factor $a$, with an a-priori arbitrary function
$g(a)$. Since holonomy operators in loop quantum gravity are related
to the discrete quantum geometry, the specific form of $g(a)$ depends
on how those structures are realized and how they reduce to the case
of isotropic models, which is referred to as lattice refinement
\cite{InhomLattice,CosConst}. At present, $g(a)$ does not appear to be
fixed uniquely by reduction schemes from the full dynamics, and so we
will parametrize it as a power law, $g(a)=g_0 a^{2x}$ with a
dimensionless number $x$ and a constant $g_0$ whose dimensions are
such that $g(a)\dot{a}$ in the holonomy exponential is
dimensionless. The dimension of $g_0$ thus depends on the value of
$x$. (Several types of phenomenological analysis, mainly in the
context of early-universe cosmology, have put restrictions on the
ranges of parameters
\cite{RefinementInflation,RefinementMatter,TensorHalfII,TensorSlowRoll,SuperInflPowerSpec}).
%
%
%%%%%%%%%%%%%%%%%%%%%%%%%%%%%%%%%%%%%%%%%%%%%%%%%%%%%%%%%%%%%%%%%%%%%%%%%%%%%%%%%%%%%%%%%%%%%%%%%%%%%%%
%
%
\subsection{Canonical variables and a harmonic system}

Holonomies are used in loop quantum gravity not at the level of
equations of motion but at the level of constraints which determine
the dynamics of phase-space points. In the present context (spatially
flat isotropic models), we have canonical variables
$\tilde{c}=\gamma\dot{a}$ and its conjugate $|\tilde{p}|=a^2$:
$\{\tilde{c},\tilde{p}\}=8\pi\gamma G/3V_0$ with the Barbero--Immirzi
parameter $\gamma$ \cite{AshVarReell,Immirzi}, the gravitational
constant $G$ and the coordinate size $V_0$ of a finite region chosen
to integrate the full symplectic structure specialized to homogeneous
fields. (Geometrically, the momentum $\tilde{p}$ actually refers to a
densitized triad and thus can take both signs, depending on the
orientation of the triad. In this article, we will assume $\tilde{p}$
to be positive without loss of generality.)  The parameter $V_0$ is
made implicit by absorbing it into the phase-space variables,
$c:=V_0^{1/3}\tilde{c}$ and $p:=V_0^{2/3}\tilde{p}$, but one still
must ensure that final results depend only on ratios of $c$ and $p$
independent of rescaling by $V_0$. See Appendix~\ref{sec:fi_vol}
for an explicit discussion of this so-called fiducial volume.

With these canonical variables, the holonomies we will use are
$\exp(if(p)c)$, where $f(p)=f_0p^x$ is obtained simply by rearranging
the previous parameters in $g(a)$: for
$f_0p^xc=f(p)c=g(a)\dot{a}=g_0a^{2x}\dot{a}$, we identify $f_0=
\gamma^{-1}V_0^{-(2x+1)/3} g_0$. With this $V_0$-dependence of $f_0$
we ensure that the product $f(p)c$ appearing in holonomies is
independent of $V_0$, in addition to being dimensionless. It will be
useful to apply a canonical transformation such that the combination
$f(p)c$ appears as a canonical variable, which must then be conjugate to
\begin{equation} \label{V}
 V:=\frac{3p}{8\pi \gamma G(1-x)f(p)}=
\frac{3a^{2(1-x)}V_0^{2(1-x)/3}}{8\pi\gamma G(1-x)f_0}
\quad\mbox{ such that }\quad  \{f_0p^xc,V\}=1\,.
\end{equation}
Since $f(p)$ scales by a factor of $V^{-1/3}$, $V\propto p/f(p)$
scales by a factor of $V_0$ if $V_0$ is changed.  The volume
of our fiducial cell, ${\rm vol}=a^3V_0$, is then obtained as:
\begin{equation}
\label{eq:V_vol}
{\rm vol} 
 = \left( \frac{8\pi\gamma G\left( 1-x\right)f_0}{3} V\right)^{\frac{3}{2(1-x)}}~.
\end{equation}
In addition to $f(p)c$ used in holonomies, another combination
independent of $V_0$ is $f(p)\sqrt{p}=f_0p^{(1+2x)/2}$. The physical
meaning of this quantity is the lattice spacing $L$ in an underlying
discrete state that gives rise to the dynamics captured in the
isotropic model. In terms of this parameter, we can write holonomies
as $\exp(i\gamma L{\cal H})$ with the Hubble parameter ${\cal H}$. The
lattice spacing can also be obtained as the ratio ${\rm vol}/V=
8\pi\gamma G(1-x)L/3$, written as
\[
 \frac{{\rm vol}}{V} = \frac{8\pi\gamma G\left( 1-x\right)}{3}
 \tilde{f}_0\tilde{p}^{\frac{1+2x}{2}}
\]
in terms of the $V_0$-independent
 $\tilde{f}_0:=f_0V_0^{(1+2x)/3}$. This equation shows that $L$ is
 independent of the scale factor only for $x=-\frac{1}{2}$.

Via holonomies, the combination $f(p)c$ appears in dynamical
equations, chiefly the Hamiltonian constraint which follows from
multiplying the Friedmann equation by $a^3$. In canonical variables,
this provides
\begin{equation}
 -\frac{1}{\gamma^2} c^2\sqrt{|p|} +\frac{8\pi G}{3}H_{\rm matter}=0
\end{equation}
with the matter Hamiltonian $H_{\rm matter}$ typically depending on
$p$ as well as on matter degrees of freedom. After the holonomy
replacement, this equation reads
\begin{equation} \label{HolFried}
 -\frac{\sin^2(f(p)c) \sqrt{|p|}}{\gamma^2 f^2(p)} +\frac{8\pi
  G}{3}H_{\rm matter}=0\,,
\end{equation}
picking the sine as a real combination to be specific. (The precise
form of this function replacing $c^2$, restricted only to be bounded
as a linear combination of $\exp(\pm i f(p)c)$ and to approach $c^2$
for small $f(p)c$ in a low-curvature regime, can be motivated from
full quantizations of curvature \cite{cosmoIII,IsoCosmo,DegFull} but
is not determined uniquely. Further quantization ambiguities arise at
this stage in addition to those in the form of $f(p)$.)

When quantized, Eq.~(\ref{HolFried}), provides the difference equation
of loop quantum cosmology. For general matter fields the resulting
behavior can be hard to analyze, but in some special cases the
dynamics becomes exactly solvable. This is the case in particular for
a free, massless scalar $\phi$ with Hamiltonian $H_{\rm matter}=
\frac{1}{2}p^{-3/2} p_{\phi}^2$ in terms of the momentum $p_{\phi}$
conjugate to $\phi$. Quantizing the system, subject to a particular
factor ordering then results in harmonic dynamics
\cite{BouncePert,BounceCohStates}, which leads to an evolution that is
not subject to quantum back-reaction; in the sense that while states
may spread and deform as they evolve, the changing shape does not
affect the motion of expectation values. Hence quantum geometry
corrections are dominant, resulting in the quantum bounce and the
boundedness results.

Harmonicity is realized thanks to a linear algebra of a complete set
of operators including the Hamiltonian in deparameterized form. We
thus first use the free scalar as an internal time variable,
formulating evolution not with respect to proper time,
so far appearing in derivatives denoted by the dot, but with respect to
$\phi$. This requires the monotonicity of $\phi (t)$, which is 
globally guaranteed in some special cases including
a constant (or vanishing) scalar potential, which are the cases
considered below. Evolution for the relational
quantities $p(\phi)$ and $c(\phi)$ is then generated by the
Hamiltonian $p_{\phi}(c,p)$ obtained as a phase-space function by
solving (\ref{HolFried}) for $p_{\phi}$ in $H_{\rm matter}$. We obtain
$p_{\phi}=\pm\sqrt{16\pi G/3}(1-x) |{\rm Im} J|$ with
\begin{equation}
 J:=V\exp(if_0p^xc)\,.
\end{equation}
The Hamiltonian is linear in $J$, and $J$ forms a linear algebra with
$V$ by taking Poisson brackets:
\begin{equation}
  \{V,J\}=-iJ\quad,\quad {}
  \{V,\bar{J}\}=i\bar{J}\quad,\quad {}
  \{J,\bar{J}\}=2iV\,.
\end{equation}
Classical equations of motion are linear and easy to solve, but even
the quantum dynamics based on the Hamiltonian $\hat{H}:=|{\rm
Im}\hat{J}|$ (dropping the constant $\sqrt{16\pi G/3}(1-x)$ for
simplicity, i.e.\ rescaling our ``time'' $\phi$) and the commutator algebra
\begin{equation}\label{comm}
  [\hat{V},\hat{J}]=\hbar\hat{J}\quad,\quad {}
  [\hat{V},\hat{J}^{\dagger}]=-\hbar\hat{J}^{\dagger}\quad,\quad {}
  [\hat{J},\hat{J}^{\dagger}]=-2\hbar(\hat{V}+\hbar/2)
\end{equation}
simplifies strongly.\footnote{Changing $V_0$ is subtle at the quantum
  level. Classically, the Poisson bracket is not preserved, implying
  that the transformation is not canonical. The
  non-canonical classical transformation then is not implemented
  unitarily. One can mimic the change of the Poisson bracket by a
  rescaling of $\hbar$, which does not amount to a physical change of
  its value but rather of the representation. Observables even at the
  quantum level do not depend on $V_0$ when these properties are
  taken into account.} For instance, equations of
  motion ${\rm d}\langle\hat{V}\rangle/{\rm d}\phi=
  \langle[\hat{V},\hat{H}]\rangle/i\hbar$ and ${\rm
  d}\langle\hat{J}\rangle/{\rm d}\phi=
  \langle[\hat{J},\hat{H}]\rangle/i\hbar$ for expectation values are
  fully determined by the expectation values and remain
  linear.\footnote{With the absolute value in $\hat{H}=|{\rm
  Im}\hat{J}|$, the Hamiltonian is not strictly linear. However, the
  Hamiltonian is preserved in the case of a $\phi$-independent
  potential, and so one only has to ensure that an initial state is
  supported solely on the positive (or negative) part of the spectrum
  of ${\rm Im}\hat{J}$ and the absolute value can be dropped for all
  the evolution. For a state required only to be semiclassical at one
  time, the spectral condition can easily be arranged. Alternatively,
  effective-constraint techniques
  \cite{EffCons,EffConsRel,EffConsComp} exist which allow one to deal
  directly with the constraint, avoiding deparameterization and taking
  square roots.} This feature, shared with the harmonic oscillator in
  quantum mechanics, implies the absence of quantum back-reaction
  i.e.\ expectation values evolve irrespective of other moments of a
  state.

%
%
%%%%%%%%%%%%%%%%%%%%%%%%%%%%%%%%%%%%%%%%%%%%%%%%%%%%%%%%%%%%%%%%%%%%%%%%%%%%%%%%%%%%%%%%%%%%%%%%%%%%%%%
%
%
\subsection{Small potential}

Harmonic models are very simple but also extremely special. For robust
conclusions, one must at least see how slight deviations from
solubility might change the conclusions derived in this simple case.
In the present context, the
main ingredient for realistic modeling is the inclusion of a
non-trivial matter potential. A number of key features change when
the scalar potential no longer vanishes. First, $\phi$ can at
best be taken as a local internal time since it may develop turning
points in a varying potential. Secondly, the Hamiltonian is no longer
in a linear algebra with basic operators, equations of motion for
expectation values no longer form a closed set, and quantum
back-reaction ensues. Dynamics can no longer be formulated in terms of
expectation values alone; we must include at least a certain number of
moments of the evolving state, which, following \cite{EffAc}, we
define as
\begin{equation}\label{Moments}
G^{\underbrace{\scriptstyle V\cdots V}_a\underbrace{\scriptstyle J\cdots J}_b}=
\langle(\hat{V}-\langle\hat{V}\rangle)^a
 (\hat{J}-\langle\hat{J}\rangle)^b\rangle_{\rm Weyl}
\end{equation}
for all positive integers $a$ and $b$ such that $a+b\geq 2$,
indicating the totally symmetric ordering of operators by the
subscript ``Weyl''. In general, all these moments are coupled
dynamically to the expectation values, providing a complicated
infinite-dimensional dynamical system. Approximately, it becomes
tractable when an initial semiclassical state is involved, whose higher
moments are suppressed by powers of $\hbar^{1/2}$ compared to lower ones. We
will make use of this approximation in our analysis.

In the presence of a non-vanishing potential $\tilde{W}(\phi)$, the expression
for $p_{\phi}$ on the constraint surface satisfying
\[
 \frac{4\pi G}{3}\frac{p_{\phi}^2}{{\rm vol}}+ {\rm vol}\tilde{W}=
 \frac{{\rm vol}^{1/3} (\Imm e^{if(p)c})^2}{\gamma^2 f(p)^2}
\]
is
\[
 p_{\phi}= \pm \sqrt{\frac{16\pi G}{3}}(1-x) \sqrt{(\Imm J)^2-
 \frac{3}{8\pi G(1-x)^2} \left(\frac{8\pi\gamma G(1-x)}{3}
 f_0V\right)^{3/(1-x)} \tilde{W}}\,.
\]
If we define
\begin{equation}\label{W}
 W(\phi):= \frac{3}{8\pi G(1-x)^2} \left(\frac{8\pi\gamma G(1-x)}{3}
 f_0\right)^{3/(1-x)} \tilde{W}
\end{equation}
and absorb the same factor in $p_{\phi}$ as in the free case, we see
that the momentum is proportional to
\begin{equation}\label{pphi}
 H(V,J):= \sqrt{({\rm Im}J)^2- V^{3/(1-x)} W(\phi)}\,,
\end{equation}
constituting a non-linear and typically ``time''-dependent
Hamiltonian. The rescaled potential $W$ depends on $V_0$ (via $f_0$)
in such a way that $V^{3/(1-x)}W$ behaves like $V_0^2$.

We will assume the potential to be always small compared to the
kinetic energy of the scalar, putting us in the kinetic-dominated
regime of loop quantum cosmology, close to the harmonic model. Sine
the harmonic model has a large class of states that remain
semiclassical (or even coherent) for long times even across the
bounce, the kinetic-dominated semiclassical regime of anharmonic
models can be expected to allow long evolution as well, an expectation
to be confirmed in this article. Larger potentials, for which states
and bounce properties can change more drastically, will require
significantly more care.

For a small potential, we can expand the square root, but terms
containing the potential remain non-linear in $V$ and $J$. Adding the
potential to the system thus provides an anharmonicity, and the exact
solubility properties of the free model disappear. Not only the
classical but also the quantum dynamics become much more complicated
and in general can be treated only by using approximations. Even if it
is possible to define the quantum dynamics in deparameterized form,
such as with a constant $W$ as used below, we have to face the
difficulty of defining a complicated square-root operator for
$p_{\phi}$.

At this stage, effective equations offer valuable tools to approximate
the quantum evolution of general semiclassical states, sidestepping
many difficulties. For a semiclassical state, we can expand the
quantum Hamiltonian
\begin{eqnarray}
H_Q(V,J,G^{a,b}) &:=&\langle H(\hat{V},\hat{J})\rangle=\langle
H(\langle\hat{V}\rangle+(\hat{V}-\langle\hat{V}\rangle),
\langle\hat{J}\rangle+(\hat{J}-\langle\hat{J}\rangle))\rangle\\
&=& H(\langle\hat{V}\rangle,\langle\hat{J}\rangle)+\sum_{a,b:a+b\geq 2} 
\frac{1}{a!b!}
\frac{\partial^{a+b}H(\langle\hat{V}\rangle,\langle\hat{J}\rangle)}{\partial
\langle\hat{V}\rangle^a\partial \langle\hat{J}\rangle^b}G^{a,b}
\end{eqnarray}
as a Hamiltonian function on the quantum phase space of states
parameterized by the expectation values $V:=\langle\hat{V}\rangle$ and
$J:=\langle\hat{J}\rangle$ together with the moments $G^{a,b}$. All
the moments, as well as the expectation values, are subject to Poisson
bracket relationships that are uniquely defined by
$\{\langle\hat{A}\rangle,\langle\hat{B}\rangle\}=
\langle[\hat{A},\hat{B}]\rangle/i\hbar$; thus, the quantum Hamiltonian
generates Hamiltonian equations of motion ${\rm d} F/{\rm
d}\phi(V,J,G^{a,b})= \{F,H_Q\}$ as usual. Also in this expansion the
quantum Hamiltonian is clearly non-linear for a non-vanishing
potential. Coupling terms between expectation values and moments imply
quantum back-reaction. For semiclassical states, high moments are
sub-dominant and the infinite sum can be truncated to finitely many
terms, starting with quantum corrections due to fluctuations and
correlations.

We note that our condition for semiclassical states is rather weak,
and so the allowed class is very general, much more general than
usually considered in evolutions of specific states. We only require
that a moment of order $a+b+1$ is suppressed by an additional power of
$\hbar^{1/2}$ compared to moments of order $a+b$. (For
dimensional reasons, the behavior must then be $G^{a,b}\sim
(V\hbar)^{(a+b)/2}$ in our variables, $V$ and $J$ having the same
dimension as $\hbar$.) This behavior is realized for Gaussian states
(which fix all moments in terms of just one or two real parameters),
but certainly not only in that case. States are allowed to spread out,
deform their shape and deviate from Gaussians. As long as the
$\hbar$ierarchy of moments is satisfied, the approximate quantum
evolution remains reliable. In quantum cosmology, the evolution
equations we use can hold true for sufficiently long times to answer
long-term questions, and their validity can be tested
self-consistently.

In the specific case at hand, the quantum Hamiltonian is
\begin{eqnarray}\label{eq:Ham}
H_Q &=& \frac{J-\bar{J}}{2i} -i\frac{V^{3/(1-x)}}{J-\bar{J}} W(\phi)-
\frac{3}{2}i\frac{2+x}{(1-x)^2} \frac{V^{(1+2x)/(1-x)}}{J-\bar{J}}
G^{VV}W(\phi) \nonumber \\
 &&+ \frac{3i}{1-x} \frac{V^{(2+x)/(1-x)}}{(J-\bar{J})^2}
  (G^{VJ}-G^{V\bar{J}}) W(\phi)\nonumber \\
&& - i\frac{V^{3/(1-x)}}{(J-\bar{J})^3}
  (G^{JJ}-2G^{J\bar{J}}+G^{\bar{J}\bar{J}}) W(\phi)
\end{eqnarray}
to second order in the moments, combined with an expansion up to
linear terms in the potential. Note that
the energy of a system of physical volume, ${\rm vol}$, is related to
$H_Q$ as 
\begin{equation} \label{eq:Energy}
 {\rm Energy} = \frac{1}{2}\frac{p_{\phi}^2}{{\rm vol}}+{\rm vol}\ \tilde{W}= 
 \frac{8\pi G}{3} (1-x)^2 H_Q^2 /{\rm vol}+{\rm
 vol}\,\tilde{W}\,,
\end{equation}
so $H_Q/{\rm vol}$ is proportional to the square-root of the {\it
kinetic energy density}.

In the following, $V$ and
$J$ refer to expectation values of the basic operators in a
state. Their products in interaction terms of the Hamiltonian thus are
not subject to factor ordering ambiguities as they would if $H_Q$ were
an operator. Also in the underlying Hamiltonian operator there are no
ordering ambiguities since the expansion comes from a Hamiltonian
quantizing (\ref{pphi}), which lacks products of non-commuting
operators. Although we have not written down an operator quantizing
the square root in (\ref{pphi}) explicitly, it is clear that the
semiclassical expansion is unaffected by ordering choices. This
statement holds provided that we choose $\hat{J}$ as one of our basic
operators together with $\hat{V}$, as is done for the free harmonic model
to make that system's Hamiltonian linear. Choosing $\hat{J}$ as the basic
variable as (opposed to, say, $f(p)c$) is consistent with our 
use of moments of $V$ and $J$.

We note that many quantum effects can be summarized as an effective
Friedmann equation \cite{QuantumBounce,BounceSqueezed}
\begin{equation}\label{EffFried}
 \left(\frac{\dot{a}}{a}\right)^2 = \frac{8\pi G}{3}\left(\rho
 \left(1-\frac{\rho_Q}{\rho_{\rm crit}}\right)
 \pm\frac{1}{2}\sqrt{1-\frac{\rho_Q}{\rho_{\rm crit}}}
\eta (\rho-P)+ \frac{(\rho-P)^2}{\rho+P}\eta^2
\right)
\end{equation}
with the critical energy density $\rho_{\rm crit}=3/8\pi G \gamma^2 L^2$ and
quantum corrections from correlations $\eta$ as well as fluctuations
entering $\rho_Q$ (and $P$ is pressure). However, quantum correlations
and fluctuations are determined by the moments of a state, and those
moments are dynamical. An effective Friedmann equation thus provides a
consistent set of differential equations only if it is accompanied by
evolution equations for the moments. In this article, we are dealing
from the outset with all equations for expectation values and moments
to a certain order, bringing all quantum degrees of freedom under control.

%
%
%%%%%%%%%%%%%%%%%%%%%%%%%%%%%%%%%%%%%%%%%%%%%%%%%%%%%%%%%%%%%%%%%%%%%%%%%%%%%%%%%%%%%%%%%%%%%%%%%%%%%%%
%
%
\subsection{Non-dynamical conditions}

In addition to the dynamics of evolving moments, further conditions
arise. First, to realize solubility of the potential-free system we
have chosen to work with a complex variable $J$, which classically
must satisfy $|J|^2=V^2$ for the original phase-space variables $c$
and $p$ to be real. Reality conditions must also be imposed at the
quantum level so that real phase-space variables are promoted to
self-adjoint operators, ensuring that one is using the correct
physical inner product to normalize states and compute expectation
values and moments. After quantization, the reality condition still
holds but is quantum corrected; we have
\begin{equation} \label{eq:reality}
 |J|^2-(V+ {\textstyle\frac{1}{2}}\hbar)^2=G^{VV}-G^{J\bar{J}}+
 \frac{1}{4}\hbar^2
\end{equation}
from taking an expectation value of the quantum reality condition
\begin{equation} \label{eq:RealOp}
 \hat{J}\hat{J}^{\dagger}=\hat{V}^2
\end{equation}
and expressing this in terms of
moments and expectation values. The reality condition thus relates
second-order moments to expectation values, but unlike the quantum
Hamiltonian it is not truncated: it is an exact relationship for a
finite number of our quantum variables. Note that in this expression
one can neglect the $\hbar^2$ terms, because we work in the
regime $V\gg \hbar$ and the second order moments are order $V\hbar$
and hence dominate the right hand side.
However for clarity we will
keep these terms and take the appropriate limits only at the end of
calculations.

In addition to (\ref{eq:reality}), which relates
expectation values in a semiclassical state with second-order moments
of order $\hbar$, reality conditions exist for the moments
themselves. By taking the expectation values of Eq.~(\ref{eq:RealOp}), one
arrives at Eq.~(\ref{eq:reality}), however one can find
further independent conditions, 
implied by taking an expectation value for instance of $\hat{V}
\hat{J} \hat{J}^\dagger = \hat{V}^3$. These new conditions relate
third-order moments 
\beq G^{VJ\bar{J}} &\equiv&
\Big\langle \left( \hat{V} - \langle \hat{V}\rangle \right)
\left( \hat{J}- \langle \hat{J}\rangle\right) \left( \hat{J}^\dagger -
\langle \hat{J}^\dagger\rangle\right) \Big\rangle_{\rm Weyl}  \\
G^{VVV} &\equiv&\Big\langle \left( \hat{V}  - \langle \hat{V}\rangle\right)^3 \Big\rangle
\eeq
to second-order ones and expectation values. Our truncation considers
the evolution equations only up to second-order moments, disregarding
third-order or higher ones, nevertheless, reality conditions obtained
at third order restrict allowed choices for second-order moments, and
thus will be important irrespective of the precise values of
third-order moments. For instance, we obtain
\be\label{eq:expression}
 G^{VJ\bar{J}} - G^{VVV} = \left( 2V +\frac{\hbar}{2}\right) G^{VV} - 
2\left( \Ree G^{VJ} \Ree J + \Imm G^{VJ} \Imm J\right)
 - \frac{\hbar}{2} \left( V^2 - \frac{\hbar}{3}V 
-\frac{\hbar^2}{3} \right)~,
\ee
an equation which contains terms of different orders in
$\hbar$. Products such as $VG^{VV}$ are of order $V^2\hbar$ in a
semiclassical state, while third-order moments behave like
$(V\hbar)^{3/2}$ and are thus smaller. 
The leading order in $\hbar$ then implies a reality
condition
\begin{eqnarray} 
 VG^{VV}&=&{\rm Re}(\bar{J}G^{VJ}) +{\cal O}\left(\hbar^{3/2}\right)~, \nonumber \\
\label{eq:RealVV}
&=&\Ree J \Ree G^{VJ} + \Imm J \Imm G^{VJ}+{\cal O}\left((V\hbar)^{3/2}\right)~,
\end{eqnarray}
for second-order moments. By analogous calculations, we obtain
\begin{equation} \label{eq:RealVJ}
 VG^{VJ}= \frac{1}{2}(JG^{J\bar{J}}+\bar{J}G^{JJ})+{\cal O}\left((V\hbar)^{3/2}\right)
\end{equation}
from the leading order of an expectation value of
$\hat{J}^2\hat{J}^{\dagger}= \hat{J}\hat{V}^2$. Eq.~(\ref{eq:RealVJ})
is a complex equation, giving two real conditions
\beq
 V\Ree G^{VJ} &=& \frac{1}{2}\left( \Ree J \Ree G^{JJ} + \Imm J \Imm G^{JJ} + \Ree J G^{J\bar{J}}
\right)+{\cal O}\left((V\hbar)^{3/2}\right)~, \nonumber \\
 V\Imm G^{VJ} &=& \frac{1}{2}\left( \Ree J \Imm G^{JJ} - \Imm J \Ree G^{JJ} + \Imm J G^{J\bar{J}}
\right)+{\cal O}\left((V\hbar)^{3/2}\right)~.
\eeq

Similarly, low-order moments arise in expectation values of products
with even more factors multiplying
$\hat{J}\hat{J}^{\dagger}-\hat{V}^2$.\footnote{Interpreting $J\bar{J}-V^2=0$
as a constraint on the classical phase space, this large set of
expectation values is an example of effective-constraint methods
developed in \cite{EffCons,EffConsRel,EffConsComp}. The reality
condition, although just one constraint, is of second class on the
non-symplectic phase space spanned by $(V,J,\bar{J})$, using
generalizations of Dirac's classification of constraints to general
Poisson manifolds \cite{brackets}. 
Thus, only conditions for the
moments result but no gauge flow need be factored out.} One could
worry that arbitrarily high orders must be considered in order to find
all reality conditions at second order. Fortunately, a combinatorial
argument shows that this is not the case, and the reality conditions
found here for second-order moments are complete. This conclusion can
also be supported by a simple counting of degrees of freedom: We
expect three independent second-order moments once all reality
conditions are implemented: two fluctuations and one correlation
parameter. Counting real and imaginary parts separately, we start with
six moments of the $(V,J)$-variables ($G^{VV}$ and $G^{J\bar{J}}$ are
always real, in contrast to $G^{VJ}$ and $G^{JJ}$). With three reality
conditions for the moments, a real one (\ref{eq:RealVV}) and one
complex one (\ref{eq:RealVJ}), the correct number results.

In addition to reality conditions, moments of a state must satisfy
uncertainty relations. In our case of a non-canonical set of basic
operators, there are three independent conditions for the second-order
moments, which can be derived by standard means:
\begin{eqnarray}\label{eq:uncert1}
 2G^{VV}\left( \Ree G^{JJ}+G^{J\bar{J}} \right)- 4(\Ree G^{VJ})^2
 &\geq& \hbar^2( \Imm J)^2\,,\\
\label{eq:uncert2}
 2G^{VV}(-\Ree G^{JJ}+ G^{J\bar{J}})- 4( \Imm G^{VJ})^2
  &\geq& \hbar^2( \Ree J)^2\,,\\
\label{eq:uncert3}
 (G^{J\bar{J}})^2-( \Ree G^{JJ})^2
- ( \Imm G^{JJ})^2
 &\geq& \hbar^2(V+\hbar/2)^2~.
\end{eqnarray}
Note that for near-saturation these equations are consistent with the
behavior $G^{a,b}\sim V\hbar$ of second-order moments with $a+b=2$.

%
%
%%%%%%%%%%%%%%%%%%%%%%%%%%%%%%%%%%%%%%%%%%%%%%%%%%%%%%%%%%%%%%%%%%%%%%%%%%%%%%%%%%%%%%%%%%%%%%%%%%%%%%%
%
%
\subsection{Evolution of the Moments}\label{sec:moments}

Following \cite{BouncePot} we obtain equations of motion for all the
moments, in addition to quantum-corrected equations for the
expectation values, by taking Poisson brackets with the quantum
Hamiltonian. These equations, even when truncated to second order in
the moments, are long, and for that reason given only in
Appendix~\ref{sec:app}, but as a finite-dimensional dynamical system
they are amenable to analytical as well as numerical investigations.
The most important one of the equations is the evolution of $V$, since
this will determine the conditions under which a bounce occurs; it is
given by
\begin{eqnarray}\label{eq:dot_V}
\frac{{\rm d}V}{{\rm d}\phi} &=& -\frac{J+\bar{J}}{2}+
\frac{J+\bar{J}}{(J-\bar{J})^2} V^{3/(1-x)} W(\phi)
- \frac{2V^{3/(1-x)}}{(J-\bar{J})^3} (G^{JJ}-G^{\bar{J}\bar{J}})W(\phi)
 \nonumber  \\
&&-\frac{6}{1-x} \frac{J+\bar{J}}{(J-\bar{J})^3}V^{(2+x)/(1-x)}
(G^{VJ}- G^{V\bar{J}})W(\phi)\nonumber \\
&&+
\frac{3}{1-x}\frac{V^{(2+x)/(1-x)}}{(J-\bar{J})^2} 
(G^{VJ}+G^{V\bar{J}})W(\phi)
\nonumber \\
&&+\frac{3}{2} \frac{2+x}{(1-x)^2} \frac{J+\bar{J}}{(J-\bar{J})^2}V^{(1+2x)/(1-x)}
G^{VV}\,W(\phi) \nonumber \\
&&+ 3 \frac{J+\bar{J}}{(J-\bar{J})^4}V^{3/(1-x)}
(G^{JJ}+G^{\bar{J}\bar{J}}-2G^{J\bar{J}})W(\phi)~.
\end{eqnarray}

In total there are nine evolution equations for the real variables
\be\label{eq:variables} \left( V, \Ree J, \Imm J, G^{VV},
G^{J\bar{J}}, \Ree G^{VJ}, \Imm G^{VJ}, \Ree G^{JJ},\Imm
G^{JJ}\right)~, \ee but not all the variables are independent owing to
the non-dynamical conditions: the reality condition (\ref{eq:reality})
combined with restrictions from the uncertainty relations. 
Moreover, (\ref{eq:RealVV}) and (\ref{eq:RealVJ}), while not
presenting sharp conditions when third-order moments are not
specified, provide ranges allowed for the moments of an initial
semiclassical state. Imposing the reality condition
(\ref{eq:reality}) leaves eight free variables. In addition, we will
require that the uncertainty conditions are nearly saturated
at least for an initial state. As a sharp condition, this
can be used to fix the initial values of, e.g., $\Ree G^{VJ}$, $\Imm
G^{VJ}$ and $G^{J\bar{J}}$, while the reality condition can be used
to eliminate (for example) $\Imm J$, thus leaving five variables: $V$,
$\Ree J$, $G^{VV}$, $\Ree G^{JJ}$ and $\Imm G^{JJ}$.  The relationship
for the deparameterized Hamiltonian, $H_Q\propto p_{\phi}$, constitutes
another equation which could be used to eliminate one further variable
in favor of $p_{\phi}$. (But note that $p_{\phi}$ is a constant of
motion only in the case of a constant potential; otherwise it is
subject to its own, potentially complicated dynamics.) The remaining
moments then constitute two fluctuation parameters, $G^{VV}$ and $\Ree
G^{JJ}$, and one correlation parameter $\Imm G^{JJ}$.\footnote{To
classify second-order quantum variables as fluctuations or
correlations, it is preferable to refer to the self-adjoint operators
$\hat{J}_{+}:=\frac{1}{2}(\hat{J}+\hat{J}^{\dagger})$ and
$\hat{J}_-:=\frac{1}{2i}(\hat{J}-\hat{J}^{\dagger})$ rather than
$\hat{J}$ and $\hat{J}^{\dagger}$. Since this is a linear
transformation, the order of moments is preserved. In particular,
$G^{JJ}= G^{J_+J_+}-G^{J_-J_-}+ 2i G^{J_+J_-}$, whose real part has
only products of the same operator, thus constituting a fluctuation,
while the imaginary part contains the product of two different
operators $\hat{J}_{\pm}$.}
For numerical evolution, we will choose initial values for these
  variables and ensure that this is compatible with (\ref{eq:RealVV})
  and (\ref{eq:RealVJ}).

Of particular interest is the evolution of the correlation parameter
$\Imm G^{JJ}$ because its value plays a role in seeing whether the
bounce of the free system generically remains if it is perturbed by
the potential. Moreover, it contributes a term to the effective
Friedmann equation (\ref{EffFried}) similar to a positive effective
potential \cite{QuantumBounce,BounceSqueezed}. Effective equations are
not yet available in complete form for a non-trivial potential
$W(\phi)$, in whose presence $\phi$ would not serve as global internal
time. Additional corrections can be expected in that case, but they
can be shown to be small for a sufficiently flat potential
\cite{BouncePot}. In our numerical discussions below
(Sec.~\ref{sec:Num}), we avoid this issue altogether by working with
constant (but non-zero) potentials, sufficiently small to ensure
kinetic domination as used in the expansions. Our goals are to provide
a more systematic analysis of consequences of potential terms compared
to what is available so far. In particular, we will analyze the
effects of a non-zero potential on the appearance of bounces and the
evolution of the moments.
%
%
%%%%%%%%%%%%%%%%%%%%%%%%%%%%%%%%%%%%%%%%%%%%%%%%%%%%%%%%%%%%%%%%%%%%%%%%%%%%%%%%%%%%%%%%%%%%%%%%%%%%%%%
%
%
\section{Analytic restrictions}

Typically one is concerned with a state that is semi-classical
initially, which in this language means that the moments are small
compared to the expectation value.  It is useful then to define the
{\it fractional} moments,
\begin{equation}
\Delta G^{VV} \equiv \frac{G^{VV}}{V^2} ~, \ \ \ \
\Delta G^{J\bar{J}} \equiv \frac{G^{J\bar{J}}}{|J|^2} ~, \ \ \ \
 \Delta G^{VJ} \equiv \frac{G^{VJ}}{V|J|}~,
 \Delta G^{JJ} \equiv \frac{G^{JJ}}{|J|^2}~,
\end{equation}
and analogously for higher-order moments $\Delta G^{a,b}$. Fractional
moments in a semiclassical state then behave as $\Delta G^{a,b}\sim
(\hbar/V)^{(a+b)/2}$.

Note that the second-order moments $\Delta G^{VV}$ and $\Delta
G^{J\bar{J}}$ (but only those) must be positive. In the free model,
$V$ and $|J|\geq {\rm Im}J\propto p_{\phi}$ take macroscopic values at
the bounce for large matter content in the kinetic-dominated regime,
all fractional moments should thus be small if an evolving state
remains semiclassical well into the bounce regime. The state remains
semiclassical at all times in the free model, so that a sufficiently
small anharmonicity brought about by the non-vanishing potential can
be expected to allow semiclassical evolution into the bounce regime
for a reasonably large class of states.

There are additional
restrictions on specific combinations of the moments, imposed by
uncertainty relations.
In terms of the fractional moments, those
relations, Eqs.~(\ref{eq:uncert1})--(\ref{eq:uncert3}), become, \beq
\label{eq:uncert1_d}
2\Delta G^{VV}\left(  \Ree \Delta G^{JJ} + \Delta G^{J\bar{J}}\right) - 4 \left( 
 \Ree\Delta  G^{VJ}\right)^2 &\geq& \hbar^2 \frac{ \left( \Imm J\right)^2}{V^2 |J|^2}~,
\\
\label{eq:uncert2_d}
2\Delta G^{VV} \left( -  \Ree\Delta  G^{JJ} + \Delta G^{J\bar{J}}\right) - 4 \left(
 \Imm\Delta G^{VJ}\right)^2 &\geq& \hbar^2 \frac{\left( \Ree  J\right)^2}{V^2|J|^2}~,
\\
\label{eq:uncert3_d}
\left( \Delta G^{J\bar{J}}\right)^2 - \left(  \Ree\Delta  G^{JJ}\right)^2
- \left(  \Imm\Delta G^{JJ}\right)^2 &\geq& \hbar^2\frac{ \left( V+\hbar/2\right)^2}{
|J|^4}~,
\eeq

Using Eq.~(\ref{eq:reality}) and Eq.~(\ref{eq:uncert3_d}) we find an inequality for
$V$,
\beq\label{eq:restriction0}
V&\geq& \frac{\hbar}{2{\cal D} \left(1+\Delta G^{VV}\right)}\Biggl[ 1 + \Delta G^{J\bar{J}}
-{\cal D} + \nonumber \\
&& \sqrt{ \left( 1 + \Delta G^{J\bar{J}} - {\cal D} \right)^2 + 2\Delta\left(
1+\Delta G^{VV}\right) \left( 1 + \Delta G^{J\bar{J}}- {\cal D}\right)}
\Biggr]~,
\eeq
where
\be\label{eq:restriction1}
{\cal D} \equiv \sqrt{ \left( \Delta G^{J\bar{J}} \right)^2 - 
\left( \Ree \Delta  G^{JJ} \right)^2 -\left(  \Imm\Delta G^{JJ}\right)^2 }~.
\ee
In addition, from
Eq.~(\ref{eq:uncert1_d}) and Eq.~(\ref{eq:uncert2_d}), we find,
\be\label{eq:restriction2}
V \geq \frac{1}{2 \sqrt{ \Delta G^{VV}
\Delta G^{J\bar{J}} -\left(\Delta \Ree G^{VJ}\right)^2 - \left( \Delta
\Imm G^{VJ}\right)^2 } }~.
\ee
These two conditions must always be
satisfied, and in particular they must be valid at the bounce, which
is by definition the minimum value of the volume. Via
Eq.~(\ref{eq:V_vol}) we find that
 the volume at the bounce is restricted. In
particular, Eq.~(\ref{eq:restriction2}) implies that,
\be
\left( \Delta G^{VV} \Delta G^{J\bar{J}} - \left(  \Ree\Delta G^{VJ}
\right)^2 -\left(  \Imm\Delta G^{VJ} \right)^2\right)\Big|_{\rm Bounce}
\rightarrow 0~,
\ee
can only occur for $V_{\rm bounce} \rightarrow \infty$. Thus a
semi-classical state (at the bounce) is
not just one in which the fractional moments are small, since the
uncertainty relations for such a state, would force the volume (of
any spatial region at the time of the
bounce) to be arbitrarily large. Second-order moments thus cannot be
self-consistently ignored.
%
%
%%%%%%%%%%%%%%%%%%%%%%%%%%%%%%%%%%%%%%%%%%%%%%%%%%%%%%%%%%%%%%%%%%%%%%%%%%%%%%%%%%%%%%%%%%%%%%%%%%%%%%%
%
%
\subsection{Zero potential}
In this section we consider, in more detail, the case of a free scalar
field, i.e.\ $W=0$, leaving the discussion of the $W\neq 0$ case to
Section~\ref{sec:W1}.
The evolution of $V$ is given by Eq.~(\ref{eq:dot_V}) and in particular
the bounce is given by $\frac{{\rm d}V}{{\rm d}\phi}=0$ which gives,
\be
\Ree  J\Big|_{\rm Bounce} = 0~.
\ee
Putting this into Eq.~(\ref{eq:reality}) we find,
\be
 \Imm J \Big|_{\rm Bounce} = \left[ \frac{V^2 \left(1+\Delta G^{VV}\right)
+\hbar\left( V + \frac{\hbar}{2}\right)}{ \left( 1 + \Delta G^{J\bar{J}}\right)} \right]^{1/2}\Big|_{\rm 
Bounce} = H_Q\Big|_{\rm Bounce}~,
\ee
where the second equality comes from Eq.~(\ref{eq:Ham}). Thus we find the density,
\be\label{eq:energy_bounce_W0}
 \frac{H_Q}{\rm vol}\Big|_{\rm Bounce} = \left[ {\rm vol}^{-1}
\sqrt{ \frac{V^2\left(1+\Delta G^{VV}\right) + \hbar\left( V + \frac{\hbar}{2}\right)}{\left( 1 + 
\Delta G^{J\bar{J}} \right)} } \right] \Big|_{\rm Bounce}~.
\ee
It is important to note that although this resembles an energy density, we have $H_Q \propto p_\phi$,
and hence the energy density is the square of Eq.~(\ref{eq:energy_bounce_W0}) (see the
comments below Eq.~(\ref{eq:Ham})).

If we assume that at the bounce, $V$ is large\footnote{The variable $V$
  depends on $V_0$ and can be made arbitrarily small (provided $x<1$)
  just by choosing a small $V_0$. But we have to be more careful at
  the quantum level due to the non-unitary implementation of changing
  $V_0$. For simplicity, we consider the limit of large $V$ as a
  substitute of the complete transformation, and ensure that physical
  values remain finite in the limit. (The limit of large $V_0$ has
  also been discussed in \cite{ShyamRev}.)},
i.e. that $V\gg \hbar$,  then the 
density at the bounce is given by the  large-$V$ limit,
which is just,
\be\label{eq:W=0}
 \frac{H_Q}{\rm vol}\Big|_{{\rm Bounce},\ {\rm large}\:V} = 
  \left[ \frac{3}{8\pi G\gamma\left(1-x\right)} \tilde{f}_0^{-1} \tilde{p}^{-\frac{1+2x}{2}} 
  \sqrt{\frac{1+\Delta G^{VV}}{1+\Delta G^{J\bar{J}}} }
   \right] \Big|_{{\rm Bounce},\ {\rm large}\:V}~,
\ee
where the fact that $\left(1+\Delta G^{VV}\right) \geq 1$ and 
$\left( 1+\Delta G^{J\bar{J}}\right) \geq 1$, ensures that the
square root is real. Note that, as expected, this is independent of
the fiducial volume $V_0$ (and
$\tilde{f}^{-1}_0\tilde{p}^{-(1+2x)/2}=L^{-1}$ is constant in our 
limit of large $V$) and so is the bounce density
\begin{equation}  \label{eq:FreeDens}
 \rho\Big|_{{\rm Bounce,\ large}\ V}= \frac{3}{8\pi
 G\gamma^2L^2}\frac{1+\Delta G^{VV}}{1+\Delta G^{J\bar{J}}}= \rho_{\rm
 crit} \frac{1+\Delta G^{VV}}{1+\Delta G^{J\bar{J}}}\,.
\end{equation}

The $\tilde{p}$-independence of the bounce density for the
specific case of $x=-1/2$ agrees with numerical results \cite{APS}, as
pointed out in \cite{QuantumBounce} whose more general results are
consistent also with the other cases of $x$. The specific dependence
of the bounce density on the moments agrees with results obtained in
the context of the effective Friedmann equation (\ref{EffFried}),
where the energy density $\rho_Q=\rho+\epsilon_0\rho_{\rm
crit}+\cdots$, according to \cite{QuantumBounce,BounceSqueezed}, has
the leading correction given by $\epsilon_0=\Delta G^{J\bar{J}}-\Delta
G^{VV}$. There is a bounce when $\rho_Q=\rho_{\rm crit}$, at which the
density is $\rho=\rho_{\rm crit}(1-\epsilon_0)$. For small fractional
moments, this is exactly (\ref{eq:FreeDens}).

Notice that even for the $x=-\frac{1}{2}$ case, the energy density at the
bounce still depends on the moments present at the bounce (even with
$W=0$, i.e.\ zero potential). The presence of these moments can
increase or decrease the energy density of the bounce compared to the
case when they are neglected, however for semi-classical states, we
can restrict the energy density to the range,
\be\label{eq:rho_range_0}
\frac{1}{\sqrt{2}} \rho_{\rm crit} < \rho\Big|_{{\rm Bounce},\ {\rm
    large}\:V} < \sqrt{2} \rho_{\rm crit}~,
\ee
where $\rho_{\rm crit}$ is the energy density of the bounce that would be
calculated neglecting the effects of the moments and we consider a
state to be semi-classical only if the factional moments are less than
unity (admittedly, a rough estimate). The cases of equality can be
removed by noting that Eq.~(\ref{eq:uncert1_d}) implies $\Delta G^{VV}
> 0$ and Eq.~(\ref{eq:uncert3_d}) gives $\Delta G^{J\bar{J}} >0$. This
is a limit on all possible semi-classical states, for which the higher
order moments can be neglected. If one wanted to consider a
more strict definition of semi-classical states (i.e.\ states for
which the fractional moments are below some specified value, less than
one), the range would be further restricted.
%
%
%%%%%%%%%%%%%%%%%%%%%%%%%%%%%%%%%%%%%%%%%%%%%%%%%%%%%%%%%%%%%%%%%%%%%%%%%%%%%%%%%%%%%%%%%%%%%%%%%%%%%%%
%
%
\subsection{Non-zero potential}\label{sec:W1}
For the more interesting case of $W \neq 0$, Eq.~(\ref{eq:dot_V}) and the
bounce condition ${\rm d}V/{\rm d}\phi=0$ gives,
\beq\label{eq:Re_Im}
 \Ree  J &=& -\frac{ W V^{\frac{3}{1-x}}}{\Imm J} \Biggl\{
\frac{1}{2} \frac{\Ree  J}{\Imm J} - \frac{3}{4} \frac{\Ree  J}{\Imm J}
\frac{|J|^2}{\left( \Imm J\right)^2} \left[  \Ree\Delta  G^{JJ} -
\Delta G^{J\bar{J}}\right] \nonumber \\
&& -\frac{3}{1-x} \frac{\Ree  J}{\Imm J} \frac{|J|}{\left( \Imm J\right)^2}
 \Imm\Delta G^{VJ} + \frac{3}{4}\frac{2+x}{\left( 1-x\right)^2} \frac{\Ree  J}{\Imm J}
\Delta G^{VV}\nonumber \\
&& -\frac{1}{2} \frac{|J|^2}{\left(\Imm J\right)^2} \Imm\Delta G^{JJ}
+ \frac{3}{2\left(1-x\right)}\frac{|J|}{\Imm J} \Ree\Delta  G^{VJ} \Biggr\}~.
\eeq

Changing to polar coordinates for the complex variable $J$ using,
\begin{equation}
 \sin  \theta  = \frac{\Imm J}{|J|}\quad,\quad
 \cos  \theta  = \frac{\Ree  J}{|J|}\quad,\quad
 \tan  \theta  = \frac{\Imm J}{\Ree  J}~,
\end{equation}
where $\theta = \arg\left(J\right)$, we can write Eq.~(\ref{eq:Re_Im}) as
\beq
 \cos \theta &=& -\frac{W V^{\frac{3}{1-x}}}{|J|^2\sin \theta}
\Bigl\{ \frac{\cos \theta}{2\sin \theta} - \frac{3\cos\theta}{4\sin^3 \theta}
\left[  \Ree\Delta  G^{JJ} - \Delta G_{J\bar{J}}\right] \nonumber \\
&&-\frac{3\cos \theta}{\left(1-x\right)\sin^2\theta}  \Imm\Delta G^{VJ}
+ \frac{3\left(2+x\right)\cos\theta}{4\left(1-x\right)^2\sin\theta}\Delta G^{VV}
\nonumber \\
&& -\frac{1}{2\sin^2\theta} \Imm\Delta G^{JJ} + \frac{3}{2\left(1-x\right)\sin
\theta} \Ree\Delta G^{VJ} \Biggr\}~.
\eeq

Now we look for solutions that are `close' to the $W=0$ case in the
near-bounce regime, by expanding around $\theta = \pi/2$ i.e.\
$\theta =  \pi/2 + \delta\theta$, which gives
$\sin \theta = 1 + {\mathcal O}\left(
\delta\theta^2\right)$ and $\cos \theta = - \delta\theta +
{\mathcal O}\left(\delta\theta^3\right)$.
We will see that this approximation is valid within the accuracy of
the expansion. With
Eq.~(\ref{eq:reality}) we get,
\[
 \delta\theta = \frac{\frac{3}{2\left(1-x\right)}
\Ree \Delta G^{VJ}   
-\frac{1}{2}  \Imm\Delta G^{JJ}}{W^{-1} V^{-\frac{3}{1-x}} \frac{ 1+\Delta
G^{J\bar{J}}}{ V^2 \left( 1+ \Delta G^{VV}\right) + \hbar\left( V + \frac{\hbar}{2}\right) }
+\frac{1}{2} - \frac{3}{4} \left( \Ree \Delta
G^{JJ} - \Delta G^{J\bar{J}} \right) - \frac{3}{1-x} \Imm \Delta
G^{VJ} + \frac{3\left( 2+x\right)}{4\left( 1-x\right)^2} \Delta G^{VV}}~. 
\]
Again, taking the
large-$V$ limit, we find
\beq\label{eq:moments_small}
\delta\theta\Big|_{{\rm large}\:V} &=& \frac{\frac{3}{2\left(1-x\right)}
\Ree \Delta G^{VJ} -\frac{1}{2} 
\Imm \Delta G^{JJ}}{
\frac{1}{2} - \frac{3}{4} \left(  \Ree\Delta
G^{JJ} - \Delta G^{J\bar{J}} \right) - \frac{3}{1-x}  \Imm\Delta
G^{VJ} + \frac{3}{4} \frac{2+x}{\left( 1-x\right)^2} \Delta G^{VV}}\,.
\eeq
Thus the expansion is valid for small fractional moments:
$\delta\theta$ is small if $ \Delta
\Imm G^{JJ}$ and $\Delta \Ree G^{VJ}$ are small. In this
approximation, there is always a solution to the bounce condition
(\ref{eq:Re_Im}).
It is clear that
the bounce cannot disappear as long as this approximation is valid. We
are thus not testing the robustness of the bounce itself, but rather
some of its properties, such as its energy density.

We can now look at the (square-root of the kinetic) energy at the
bounce which from (\ref{eq:Ham}) is given by,
\beq
H_Q\Big|_{\rm Bounce}
&=& |J| \sin \theta - \frac{W}{|J|\sin\theta} V^{\frac{3}{1-x}}
\nonumber \\
&&\times\left\{ \frac{1}{2} +\frac{3\left( 2+x\right)}{4\left(1-x\right)^2} 
\Delta G^{VV} - \frac{3 \Imm \Delta G^{VJ}}{2\left(1-x\right)\sin\theta}
- \frac{1}{4\sin^2\theta} \left( \Ree \Delta G^{JJ} - \Delta G^{J\bar{J}}\right)
\right\}~.\nonumber \\
\eeq
Once again, expanding about $\theta = \pi/2 + \delta \theta$,
gives
\beq\label{eq:Ham_W1}
H_Q\Big|_{\rm Bounce} &=&  |J| - \frac{W V^{\frac{3}{1-x}}}{|J|}
\Biggl\{ \frac{1}{2} + \frac{3\left( 2+x\right)}{4\left( 1-x\right)^2}
\Delta G^{VV} - \frac{3 \Imm \Delta G^{VJ}}{2\left( 1-x\right)} \nonumber \\
&&- \frac{1}{4}\left( \Ree \Delta G^{JJ} - \Delta G^{J\bar{J}} \right)
 + {\mathcal O}\left( \delta \theta\right)^2 \Biggr\}~.
\eeq
Here it is important to remember our redefinition of the potential:
every $W\left(\phi\right)$ comes with a $V_0$-dependent factor via
$f_0$ in (\ref{W}).
Using this and the reality condition Eq.~(\ref{eq:reality}) one finds that
the (square-root of kinetic) energy density at the bounce, again
taking the large-$V$ limit, is given by,
\beq\label{eq:Ham_W1_approx}
 \left(\frac{H_Q}{\rm vol} \right)\Big|_{{\rm Bounce},\ {\rm large}\:V}
&=& A - \tilde{W} A \left(\frac{1+\Delta G^{J\bar{J}}}{1+\Delta G^{VV}}\right)
\left( \frac{ 8\pi \gamma G \tilde{f}_0^2 \tilde{p}^{1+2x} }{3} \right)
 \nonumber \\
&&\Biggl[ \frac{1}{2} + \frac{3\left(2+x\right)}{4\left(1-x\right)^2}
\Delta G^{VV} - \frac{3 \Imm \Delta G^{VJ}}{2\left( 1-x\right)}
 - \frac{1}{4}\left( \Ree \Delta  G^{JJ} - \Delta G^{J\bar{J}}\right)
+ {\mathcal O}
\left(\delta\theta\right)^2\Biggr]~,
\eeq
where
\be
 A \equiv  \left\{ \frac{|J|}{\rm vol} \right\}_{{\rm large}\:V}~.
\ee
Using the reality condition Eq.~(\ref{eq:reality}) and Eq.~(\ref{eq:V_vol}) 
we see that this is,
\be
 A = \frac{3}{8\pi G\gamma\left( 1-x\right)} \tilde{f}_0^{-1} \tilde{p}^{-\frac{1+2x}{2}} 
        \sqrt{ \frac{1+\Delta G^{VV}}{1+ \Delta G^{J\bar{J}}} } = r(0)~,
\ee
which is just the (square-root of the) energy density at the bounce that we had for the $W=0$ case,
Eq.~(\ref{eq:energy_bounce_W0}). So finally we have that
while the bounce is `close' to
the $W=0$ case, its (square-root of the kinetic) energy density is given by,
\beq\label{eq:ED_bounce_2}
 r\left( W\right) &=& r\left(0\right) - \tilde{W}r\left(0\right) 
\left( \frac{ 8\pi \gamma G \tilde{f}_0^2 \tilde{p}^{1+2x} }{3} \right)
 \left( \frac{1+\Delta G^{J\bar{J}}}{1+\Delta G^{VV} }\right)
\nonumber \\
&&
 \left[ \frac{1}{2} + \frac{3\left(2+x\right)
}{4\left(1-x\right)^2}\Delta G^{VV} - \frac{3}{2\left( 1-x\right)} \Imm  \Delta
G^{VJ} + \frac{1}{4}\left( \Ree \Delta G^{JJ} - \Delta G^{J\bar{J}}
\right) \right] + {\mathcal O}\left(\delta\theta\right)^2~,
\eeq
where
\be
 r\left(W\right) = \frac{H_Q}{\rm vol}\Big|_{{\rm Bounce},\ {\rm large}\:V}.
\ee
Just as in the $W=0$ case this is explicitly independent of the fiducial volume.

For the $x=-1/2$ case, for instance, this is
\be\label{eq:bounce_ed_W}
 r\left(W\right) = r\left(0\right) - \tilde{W} r\left(0\right)\left( \frac{8\pi G\gamma\tilde{f}_0^2}{3}
\right)
\left( \frac{ 1+ \Delta G^{J\bar{J}}}{1+\Delta G^{VV}}\right)\left[ \frac{1}{2} + \frac{1}{2}
 \Delta G^{VV} - \Imm \Delta G^{VJ} - \frac{1}{4}\left( \Ree \Delta G^{JJ} - \Delta G^{J\bar{J}}
\right) \right] + {\mathcal O}\left(\delta\theta\right)^2~,
\ee
and the small parameter that we have expanded in is,
\beq
 \delta\theta\Big|_{{\rm large}\:V} &=& \frac{\Delta \Ree  G^{VJ}-\frac{1}{2}  \Imm\Delta
G^{JJ} }{\frac{1}{2} - 
\frac{3}{4} \left( \Delta \Ree  G^{JJ} 
- \Delta G^{J\bar{J}} \right) - 2  \Imm \Delta G^{VJ}
 + \frac{1}{2} \Delta G^{VV}}\,.
\nonumber
\eeq

In general, we can say that the (square-root of the kinetic) energy
density at the bounce for $x=-1/2$, for semi-classical states (i.e.\
states with fractional moments less than unity), is restricted to the range,
\be
\sqrt{2}  r(0)
\left[ 1 - \left( \frac{ 8\pi G\gamma \tilde{f}_0^2}{3}\right)
 \tilde{W} \left(\frac{1+\Delta G^{J\bar{J}}}{1+\Delta G^{VV}}\right) I\right]
\geq r\left(W\right) \geq
\frac{r(0)}{\sqrt{2}}\left[ 1 - \left( \frac{8\pi G\tilde{f}_0^2}{3}\right)\tilde{W}\left(
\frac{1+\Delta G^{J\bar{J}}}{1+\Delta G^{VV}}\right) I\right]
\ee
with
\be\label{eq:sign1}
I\equiv  \frac{1}{2} +  \frac{1}{2}  \Delta G^{VV} - \Imm \Delta 
G^{VJ} - \frac{1}{4}\left( \Ree  \Delta G^{JJ} - \Delta G^{J\bar{J}}
\right) ~,
\ee
where we used Eq.~(\ref{eq:rho_range_0}) and, as before, $r(0)$ is
the (square-root of the kinetic) energy
density that would have been calculated neglecting the moments entirely.

To decide whether the effect of moments raises or lowers bounce
densities, it is more convenient to consider the total rather than
kinetic energy density. After all, from the effective Friedmann
equation (\ref{EffFried}) we expect the total density to trigger
bounces. With (\ref{eq:Energy}), we have the energy density at the
bounce given by
\begin{equation}\label{eq:energy_den}
 \rho_{\rm Bounce}= \frac{8\pi G}{3}(1-x)^2 \left(\frac{H_Q}{\rm
 vol}\right)^2+ \tilde{W}= \rho_{\rm crit}\frac{1+\Delta G^{VV}}{1+\Delta G^{J\bar{J}}}- D\tilde{W}
\end{equation}
with the critical density $\rho_{\rm crit}$
expected for the free bounce, and (for $x=-1/2$)
\be\label{eq:sign2}
D=2I-1=
\Delta G^{VV} - 2 \Imm \Delta
G^{VJ} - \frac{1}{2}\left( \Ree \Delta G^{JJ} - \Delta G^{J\bar{J}}
\right)\,.
\ee
Also this correction to the bounce density agrees with the term
$\delta_1$ from \cite{QuantumBounce} (or $\epsilon_1$ from
\cite{BounceSqueezed}) based on 
the effective Friedmann
equation. By Eq.~(\ref{eq:uncert2_d}) we have
\be\label{eq:inequal1}
   -(\Ree\Delta  G^{JJ} - \Delta G^{J\bar{J}}) \geq 0
\ee
and $\Delta G^{VV}$ is positive while $\Delta\Imm G^{VJ}$ does not
have definite sign. The sign of $D$ can thus take both values, raising
or lowering the bounce density on top of the effect due to the potential.%
%PRD1:
\footnote{In \cite{DensityOp}, it was shown that expectation values of
a density operator, of the form $\langle{\rm Energy}/{\rm
volume}\rangle$, are bounded from below by the critical density of the
harmonic model even in the presence of a cosmological constant. Our
statement is not in conflict with this result because we are
considering bounce values for expressions of the form $\langle{\rm
Energy}\rangle/\langle{\rm volume}\rangle$. Obviously, the two
expressions differ by terms depending on the state via its moments,
exactly the kind of terms we compute; what we show is that those
moments may pull the bounce density below the critical one of the
harmonic model. In this context, one should note that \cite{DensityOp}
refers to a {\em density operator} which cannot exist in full loop
quantum gravity where only total Hamiltonians, but not energy
densities of matter or the gravitational field can be represented as
well-defined operators. Expressions such as $\langle{\rm
Energy}\rangle/\langle{\rm volume}\rangle$ used here, on the other
hand, have completely well-defined analogs in the full theory. Thus,
our results have a higher degree of robustness than those of
\cite{DensityOp}, even though the mathematical statements of
\cite{DensityOp} are sharper.}
%
%
%%%%%%%%%%%%%%%%%%%%%%%%%%%%%%%%%%%%%%%%%%%%%%%%%%%%%%%%%%%%%%%%%%%%%%%%%%%%%%%%%%%%%%%%%%%%%%%%%%%%%%%
%
%
\section{Dynamical coherent states}\label{sec:dyn_coherence}
Of particular interest are states which saturate the uncertainty bounds Eqs.~(\ref{eq:uncert1})--(\ref{eq:uncert3})
and remain saturating them throughout the evolution of the system. Such states are referred to
as {\it dynamical coherent states}. We will first construct such states for the $W\left(\phi\right)=0$
case which is analytically tractable and then discuss the consequences for the rather more
complicated case of $W\left(\phi\right)\neq 0$.

For a state to initially saturate the uncertainty conditions we require,
\begin{eqnarray}
\label{eq:uncert_s_1}
2G^{VV}\left( \Ree G^{JJ}+G^{J\bar{J}} \right)- 4(\Ree G^{VJ})^2
 - \hbar^2( \Imm J)^2 &=& 0~, \\
\label{eq:uncert_s_2}
2G^{VV}(-\Ree G^{JJ}+ G^{J\bar{J}})- 4( \Imm G^{VJ})^2
  - \hbar^2( \Ree J)^2 &=& 0~,\\
\label{eq:uncert_s_3}
 (G^{J\bar{J}})^2-( \Ree G^{JJ})^2
- ( \Imm G^{JJ})^2
 - \hbar^2(V+\hbar/2)^2 &=& 0~.
\end{eqnarray}
These states must also satisfy the reality condition,
Eq.~(\ref{eq:reality}), which we rewrite here for convenience,
\be\label{eq:reality2}
|J|^2-(V+ {\textstyle\frac{1}{2}}\hbar)^2=G^{VV}-G^{J\bar{J}}+
\frac{1}{4}\hbar^2~. 
\ee
Dynamical coherent states (as with any other state), must also 
satisfy Eqs.~(\ref{eq:RealVV})--(\ref{eq:RealVJ}), however these are
`weak' constraints in situations in which third-order moments are not
specified, in the sense that they need only be satisfied up to
order $(V\hbar)^{3/2}$ and hence do not provide sharp constraints on 
the states.

A state that satisfies these four conditions will
initially saturate the uncertainty bounds, however to ensure that it
remains saturating them we also require that the derivatives of
Eqs.~(\ref{eq:uncert_s_1})--(\ref{eq:uncert_s_3}) with respect to $\phi$
be zero. For the $W\left(\phi\right)=0$ case one can easily show that
the derivative of Eq.~(\ref{eq:uncert_s_1}) with respect to $\phi$
automatically vanishes, while for Eq.~(\ref{eq:uncert_s_2}) and
Eq.~(\ref{eq:uncert_s_3}) we require,
\be\label{eq:cond_4} 2
\Ree G^{VJ} \left[ \Ree G^{JJ} - G^{J\bar{J}} \right] + 2 \Imm G^{VJ} \Imm
G^{JJ} + \hbar^2\left( V+\frac{\hbar}{2}\right)\Ree J =0~.
\ee
This one constraint is sufficient to ensure that both
Eq.~(\ref{eq:uncert_s_2}) and Eq.~(\ref{eq:uncert_s_3}) are
independent of $\phi$. This makes the space of dynamical coherent
states a hypersurface of co-dimension one in the space of states that
instantaneously saturate the uncertainty conditions. Explicit
solutions have been found and discussed in
\cite{BounceCohStates,Harmonic}. The existence of a large class of
coherent states is analogous to the well-known feature of the harmonic
oscillator in quantum mechanics, but notice that potentially
significant spreading of states is generically possible even for
dynamical coherent states \cite{BeforeBB,Harmonic}.

The full parameter space of this system is $9$-dimensional, however the five
constraints above can be solved to give dynamical coherent states in a  
4-dimensional parameter sub-space. A particularly convenient way of solving these
constraints is to specify initial data
\be
\left( V,\ \Imm J,\ \Ree G^{JJ},\ G^{J\bar{J}}\right)~,
\ee
solve Eqs.~(\ref{eq:uncert_s_1})--(\ref{eq:uncert_s_3}) as,
\beq
 \Ree G^{VJ} &=& \pm \sqrt{ \frac{1}{2}G^{VV}\left( \Ree G^{JJ}+G^{J\bar{J}}
  \right) - \frac{1}{4}\hbar^2( \Imm J)^2 }~, \\
  \Imm G^{VJ} &=& \pm \sqrt{ \frac{1}{2}G^{VV}(-\Ree G^{JJ}+ G^{J\bar{J}})
  - \frac{1}{4}\hbar^2( \Ree J)^2 = 0}~, \\
  \Imm G^{JJ} &=& \pm \sqrt{ (G^{J\bar{J}})^2-( \Ree G^{JJ})^2
  - \hbar^2(V+\hbar/2)^2}~,
\eeq
and then simultaneously solve Eq.~(\ref{eq:reality2}) and Eq.~(\ref{eq:cond_4}) for
$G^{VV}$ and $\Ree J$. There is an analytic solution to this system, however in
practice it is more useful to solve the system for some given initial data i.e.\
to specify $\left( V,\ \Imm J,\ \Ree G^{JJ},\ G^{J\bar{J}}\right)$ and solve the
resulting simultaneous equations.
This is a useful test of the numerical implementation of the system, which we
discuss in the following section.

In order to attempt a construction of dynamical coherent states for the $W\left(\phi\right)\neq 0$
case, one can in principle follow the same procedure. Clearly the initial states
will still have to satisfy the constraints given by Eqs.~(\ref{eq:uncert_s_1})--(\ref{eq:reality2}),
however in order for these conditions to be independent of $\phi$ we now need to 
additionally satisfy,
\beq\label{eq:cond_4_W_1}
2\frac{{\rm d}G^{VV}}{{\rm d}\phi}\left[ \Ree G^{JJ} + G^{J\bar{J}}\right]
+2G^{VV}\left[ \frac{{\rm d}\Ree G^{JJ}}{{\rm d} \phi} + 
\frac{{\rm d}G^{J\bar{J}} }{{\rm d}\phi}
\right] - 8 \Ree G^{VJ} \frac{{\rm d} \Ree G^{VJ}}{{\rm d} \phi}
-2\hbar^2\Imm J \frac{{\rm d} \Imm J}{{\rm d}\phi } =0~, \\ 
\label{eq:cond_4_W_2}
2\frac{{\rm d}G^{VV}}{{\rm d}\phi}\left[ - \Ree G^{JJ} + G^{J\bar{J}}\right]
+2G^{VV}\left[ - \frac{{\rm d}\Ree G^{JJ}}{{\rm d} \phi} + 
\frac{{\rm d}G^{J\bar{J}} }{{\rm d}\phi}
\right] - 8 \Imm G^{VJ} \frac{{\rm d} \Ree G^{VJ}}{{\rm d} \phi}
-2\hbar^2\Ree J \frac{{\rm d} \Ree J}{{\rm d}\phi } =0~, \\
\label{eq:cond_4_W_3}
2G^{J\bar{J}}\frac{{\rm d}G^{J\bar{J}}}{{\rm d}\phi} - 2 \Imm G^{JJ}
\frac{{\rm d}\Imm G^{JJ}}{{\rm d}\phi} - 2\hbar^2 \left( V+\frac{\hbar}{2}\right)
\frac{{\rm d}V}{{\rm d}\phi} = 0~,
\eeq
where the derivatives of $G^{a,b}$ are given in Appendix~\ref{sec:app}. These three equations
need not be degenerate and certainly are not automatically zero for general states
and general $W\left(\phi\right)$. They correspond to the single equation
Eq.~(\ref{eq:cond_4}) in the $W\left(\phi\right)=0$ case. These equations are
significantly more difficult to solve. Naively it would appear that 
we now have seven constraints on the nine-dimensional parameter space that give dynamical
coherent states for $W\left(\phi\right)\neq 0$. In fact it is easy to see that
Eqs.~(\ref{eq:cond_4_W_1})--(\ref{eq:cond_4_W_3}) contain at least four independent
conditions on the moments. To see this note that all of the derivatives
given in Appendix~\ref{sec:app} are of the form,
\be
 \frac{{\rm d} F}{{\rm d}\phi} = F_0 + W\left(\phi\right) F_1~,
\ee
where $F_0$ and $F_1$ have no dependence on $W\left(\phi\right)$. It is then clear that
Eqs.~(\ref{eq:cond_4_W_1})--(\ref{eq:cond_4_W_3}) contain terms that are independent of
$W\left(\phi\right)$ and those that are not. For a general potential, these must independently cancel which gives
us four conditions. The $W\left(\phi\right)$ independent part of Eqs.~(\ref{eq:cond_4_W_1})--(\ref{eq:cond_4_W_3})
gives exactly Eq.~(\ref{eq:cond_4}), while the $W\left(\phi\right)$ dependent parts
give three additional constraints. 
Thus we are left with (at most) a one dimensional parameter
space that can support dynamical coherent states for a general $W\left(\phi\right)$.

Thus far we have been concerned only with the conditions imposed on the moments
by the requirement that the uncertainty conditions be saturated and their first
derivatives be zero. This alone is not sufficient to fix a dynamical coherent
state, since we have to require that all derivatives of the uncertainty conditions 
vanish. For the $W\left(\phi\right)=0$ case, one can explicitly check that
the derivative, with respect to $\phi$, of Eq.~(\ref{eq:cond_4}) vanishes
provided the uncertainty conditions are saturated (in particular provided
Eq.~(\ref{eq:uncert_s_2}) and Eq.~(\ref{eq:uncert_s_3}) are saturated,
since Eq.~(\ref{eq:uncert_s_1}) decouples from the
system in this case). Thus by inspection Eq.~(\ref{eq:cond_4}) is the only
non-trivial condition that needs to be satisfied.

For the $W\left(\phi\right) \neq 0$ case we have four constraints to investigate.
The $W\left(\phi\right)$-independent part of Eqs.~(\ref{eq:cond_4_W_1})--(\ref{eq:cond_4_W_3})
is just Eq.~(\ref{eq:cond_4}) and hence the previous argument holds.
For the $W\left(\phi\right)$ dependent parts of Eqs.~(\ref{eq:cond_4_W_1})--(\ref{eq:cond_4_W_3})
the situation is less clear, due to the complexity of the evolution equations
given in Appendix~\ref{sec:app}. However if any one of the derivatives of any of
$W\left(\phi\right)$ dependent parts of Eqs.~(\ref{eq:cond_4_W_1})--(\ref{eq:cond_4_W_3})
does not vanish then the parameter space that supports dynamical coherent states
reduces to a single point and if any two of the derivatives fail to vanish
dynamical coherent states cannot exist.

In addition to the fact that dynamical coherent state may not exist,
this reduction in the dimension of the parameter space for such states
is also a difficulty for study of the dynamics. Since one can no
longer explicitly solve the eight constraints implied by
Eqs.~(\ref{eq:uncert_s_1})--(\ref{eq:reality2}) and
Eqs.~(\ref{eq:cond_4_W_1})--(\ref{eq:cond_4_W_3})\footnote{Recall
Eqs.~(\ref{eq:cond_4_W_1})--(\ref{eq:cond_4_W_3}) are four independent equations.},
small changes in
the initial state can lead to radically different trajectories.  This
is an inherent property of the evolution equations of the states and
can thus play a significant role in both analytic and numerical
investigations of the dynamics of (nearly) dynamical coherent states
(if such states exist). Numerically, difficulties can arise from the
large parameter space to be explored when one can no longer stay on
the saturation surface. One may reduce the number of initial
conditions by starting on the saturation surface, but numerical errors
easily lead one into danger of falling onto the wrong side of
saturation.
%
%
%%%%%%%%%%%%%%%%%%%%%%%%%%%%%%%%%%%%%%%%%%%%%%%%%%%%%%%%%%%%%%%%%%%%%%%%%%%%%%%%%%%%%%%%%%%%%%%%%%%%%%%
%
%
\section{Numerical evolution of the system}
\label{sec:Num}

The evolution of the moments and their back-reaction on the trajectories of $\left( J, V\right)$ 
can be directly implemented numerically. In particular, because Eq.~(\ref{eq:dot_V}) and the equations
in Appendix~\ref{sec:app} are flows in configuration space (i.e.\ depend explicitly only on the 
variables and not their derivatives) the numerical implementation is rather straightforward.
Care is however needed to ensure that our numerics are sufficiently accurate so as not to introduce
violations of the uncertainty relations, Eqs.~(\ref{eq:uncert1})--(\ref{eq:uncert3}),
or reality constraints, Eq.~(\ref{eq:reality}), (\ref{eq:RealVV}) and (\ref{eq:RealVJ}), when exact
conservation is expected
analytically. In the $W \neq 0$ case, as already discussed, exact conservation is not expected since
the absence of higher order moments are expected to introduce a fundamental limit to the accuracy of our
evolution. Nevertheless, as long as the fractional moments are small, all conditions should be respected.

Another possible point of concern for the general $W\left(\phi\right)\neq 0$ case, is that it is no longer guaranteed
that $\phi$ will be monotonic. If it is not, then the deparametrization of the system (using
$\phi$ as an internal time) will no longer be valid. Such a break down would correspond to
a non-unitary evolution of the system and is clearly not physical. This issue is easily
avoided, by choosing the $\phi$ coordinate only in regions where it is monotonic and
changing coordinates at turn-around points. In our numerical investigation, we avoid this
subtlety entirely by focusing our attention on the $W\left(\phi\right)={\rm const}\not=0$ cases,
for which $\phi$ {\it is} monotonic throughout the evolution. Here, we choose $W<0$ so as to get
into a bounce regime at larger densities and smaller volume than in the free case.

In practice, fractional moments grow, which can result in the
non-conservation of the uncertainty conditions and limiting the length
of $\phi$ for which the numerical equations can be evolved. However
this is sufficient to examine the bounce, for suitable chosen
states.
To monitor the growth of higher-order moments which are not included
in the evolution equations, we will use plots of the higher-order
terms in reality conditions:
\begin{eqnarray} 
 VG^{VV}-{\rm Re}(\bar{J}G^{VJ})&=:& V^3 R_1\,,\label{eq:R1}\\
 V\Ree G^{VJ} - \frac{1}{2}\left( \Ree J \Ree G^{JJ} + \Imm J \Imm G^{JJ} + \Ree J G^{J\bar{J}}
\right) &=:& V^3 R_2\,,\\
 V\Imm G^{VJ} - \frac{1}{2}\left( \Ree J \Imm G^{JJ} - \Imm J \Ree G^{JJ} + \Imm J G^{J\bar{J}}
\right) &=:& V^3 R_3\,.
\end{eqnarray}
>From the general behavior of the moments, we expect $R_1$, $R_2$ and
$R_3$ to be significantly smaller than the fractional second-order
moments, e.g.\ $R_1\sim (\Delta G^{VV})^{3/2}\sim (\hbar/V)^{3/2}$.
In addition, it is rather difficult to find initial conditions
that simultaneously satisfy all of the uncertainty conditions. Thus it
is often more convenient to use the uncertainty conditions to fix some
of the initial data, by requiring that the state {\it under saturate}
Eqs.~(\ref{eq:uncert1})--(\ref{eq:uncert3}) by some constant i.e.\
that the state satisfy, for example,
\be\label{eq:uncert_under}
2\Delta G^{VV}\left( \Delta \Ree G^{JJ} + \Delta G^{J\bar{J}}\right) - 4 \left( 
\Delta \Ree G^{VJ}\right)^2 = \hbar^2 \frac{ \left( \Imm J\right)^2}{V^2 |J|^2} + U_1~,
\ee
and similarly for the remaining two uncertainty conditions. In general $U_1$, $U_2$ and 
$U_3$ depend on $\phi$, however, from Section~\ref{sec:dyn_coherence}
we see that for the $W=0$ case, $U_1$ should be a constant, while $U_2$ and $U_3$
should evolve similarly.

\begin{figure}
 \begin{center}
  \includegraphics{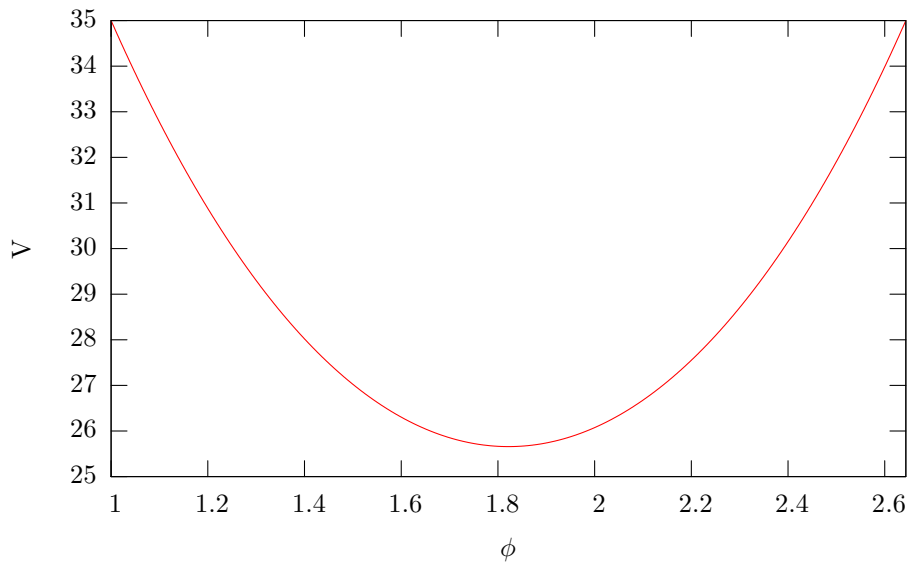}
  \caption{ \label{fig:1} The trajectory $V\left(\phi\right)$, evolved numerically for the initial conditions
given by Eq.~(\ref{eq:ini_conditions}), for the zero potential ($W\left(\phi\right)=0$) case. 
Note that only the region around the bounce is plotted. This bounce occurs at $V\approx 25.6582$ with a corresponding
(square-root of the) energy density of $0.9743$. Note that for this trajectory, the bounce occurs at
$\phi_{\rm Bounce} \approx 1.821815$.
}
 \end{center}
\end{figure}

For the numerical calculations we used units in which $\hbar = 1$ and $V$ is dimensionless.
A typical example for $W\left(\phi\right)=0$ is given in Figure~\ref{fig:1}, which had the following
initial conditions:
\be\label{eq:ini_conditions}
\left( \begin{array}{ccccccccc}
 V & \Ree J & \Imm J & \Delta G^{VV} & \Delta G^{J\bar{J}} & \Delta \Ree G^{VJ} &  \Imm\Delta G^{VJ}
&  \Ree\Delta G^{JJ} &  \Imm\Delta G^{JJ} \\
35.0 & 24.0 & 25.0 & 0.0331 & 0.0833 & 0.0377 & 0.0294 & 0.0250 & 0.0737
\end{array}\right)~,
\ee
with $x=-1/2$. These initial conditions 
under-saturate the uncertainty conditions, Eq.~(\ref{eq:uncert1})--(\ref{eq:uncert3}) 
and also saturates their derivatives with respect to $\phi$ (Eq.~(\ref{eq:cond_4})).
In addition, one can explicitly check that Eqs.~(\ref{eq:RealVV})--(\ref{eq:RealVJ})
are satisfied, up to the correct order, throughout the evolution of the system. In the 
numerical evolutions, we have taken $\hbar=1$, hence one needs to check that 
Eqs.~(\ref{eq:RealVV})--(\ref{eq:RealVJ}) are satisfied up to order $\sqrt{\Delta G_{VV}}$,
(or other second order moments), since this is the order of the terms neglected.

For this state the bounce occurs at $V\left(\phi_{\rm Bounce}\right)
= 25.6582$, with the square-root of the dimensionless kinetic energy density $r\left(0\right) = 0.974348$.
This can be compared
to the expected value, given by Eq.~(\ref{eq:energy_bounce_W0}), of $r\left(0\right) = 0.974349$,
showing the excellent agreement of the numerical implementation with the analytic solutions.
In the $W\left(\phi\right)=0$ case, the evolution equations given in Appendix~\ref{sec:app}
are exact and hence we can analytically calculate the (square-root of the) energy density at the bounce to
arbitrary accuracy (for any given value of $\Delta G_{VV}$ and $\Delta G_{J\bar{J}}$ at the 
bounce). This is not true for $W\left(\phi\right) \neq 0$, since the evolution equations
are valid only up to order ${\mathcal O}\left( W^2\right)$. However even for the $W=0$ case
numerical approximation of the derivatives introduces an error, this can become significant
for states that (almost) exactly saturate the uncertainty bounds, where numerical
artifacts may lead to unphysical trajectories.

\begin{figure}
 \begin{center}
  \includegraphics{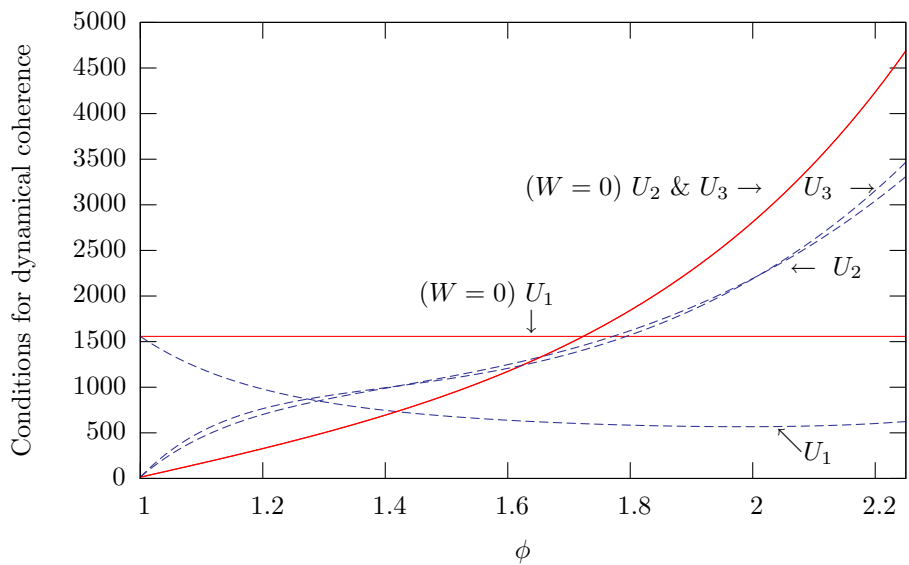}
  \caption{ \label{fig:1.75} 
The system was initialized with a state that both under-saturates the uncertainty
conditions, Eq~(\ref{eq:uncert1})--(\ref{eq:uncert3}) and ensures that the derivatives with respect to
$\phi$ are all positive.
Plotted are the values of $U_1$, $U_2$, $4.0\times U_3$  for the $W\left(\phi\right)=0$ and $W
\left(\phi\right)=-0.05$ cases. Recall that $U_1$, $U_2$ and $U_3$ positive indicates that the 
uncertainty conditions are {\it under saturated} i.e.\ physically acceptable. Clearly all the
uncertainty conditions are satisfied, in particular at the bounce, which occurs at $\phi \approx 1.8$.
Notice that for $W=0$, $U_1$ is essentially constant as expected from the discussion above
Eq.~(\ref{eq:cond_4}). Note that the initial conditions are deliberately taken so as to ensure
that the system is far from the uncertainty conditions throughout the bounce.}
 \end{center}
\end{figure}

If the saturation of the uncertainty conditions, Eqs.~(\ref{eq:uncert_s_1})--(\ref{eq:uncert_s_3}),
and their derivatives (Eq.~(\ref{eq:cond_4}) for the $W\left(\phi\right)=0$ case) were met exactly then the
state would saturate the uncertainty conditions and remain saturating them, through out the
evolution in $\phi$. For the state here, the initial data was chosen so that the uncertainty
conditions, Eq.~(\ref{eq:uncert_under}), were {\it initially} under-saturated and the
the derivatives of these equations with respect to $\phi$ are positive for both the $W\left(\phi\right)=0$
and $W\left(\phi\right)=-0.05$ cases. This ensures that, at least initially, the system is
evolving towards the under-saturated side of the uncertainty condition hypersurface. Figure~\ref{fig:1.75}
shows how the uncertainty constraints evolve for both the $W=0$ and $W=-0.05$ cases.

\begin{figure}
 \begin{center}
  \includegraphics{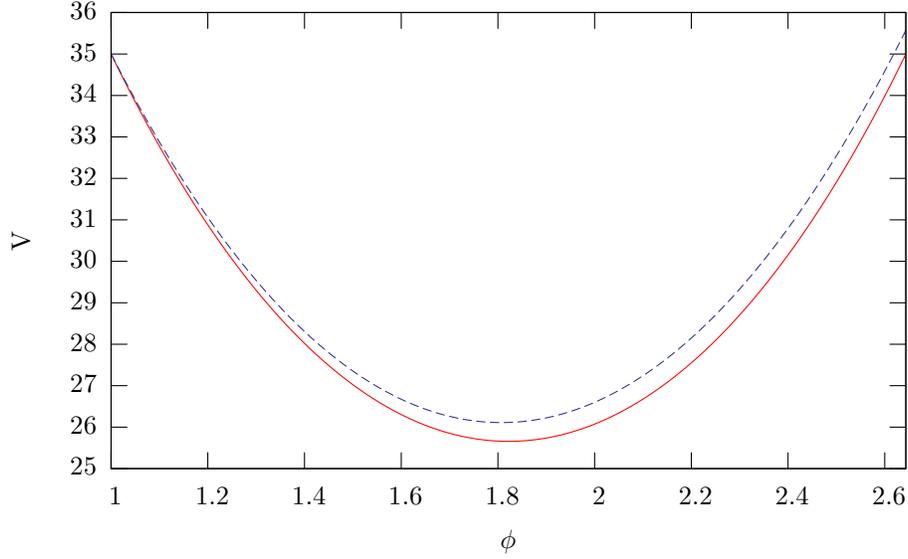}
  \caption{ \label{fig:3} The trajectory $V\left(\phi\right)$, evolved numerically for the initial conditions
given by Eq.~(\ref{eq:ini_conditions}). Again only the bounce region is plotted. The solid curve correspond to the
$W\left(\phi\right)=0$ case and the dashed curve to the $W\left(\phi\right)=-0.05$ case. The bounce for this
latter case occurs at $V\approx 25.11053$ with a corresponding square-root energy density of $1.00515$.
This is, approximately $3.2\%$ higher than the $W\left(\phi\right)=0$ case, a change mainly (but not solely)
attributed to the negative potential energy; see below. }
 \end{center}
\end{figure}

\begin{figure}
 \begin{center}
  \includegraphics{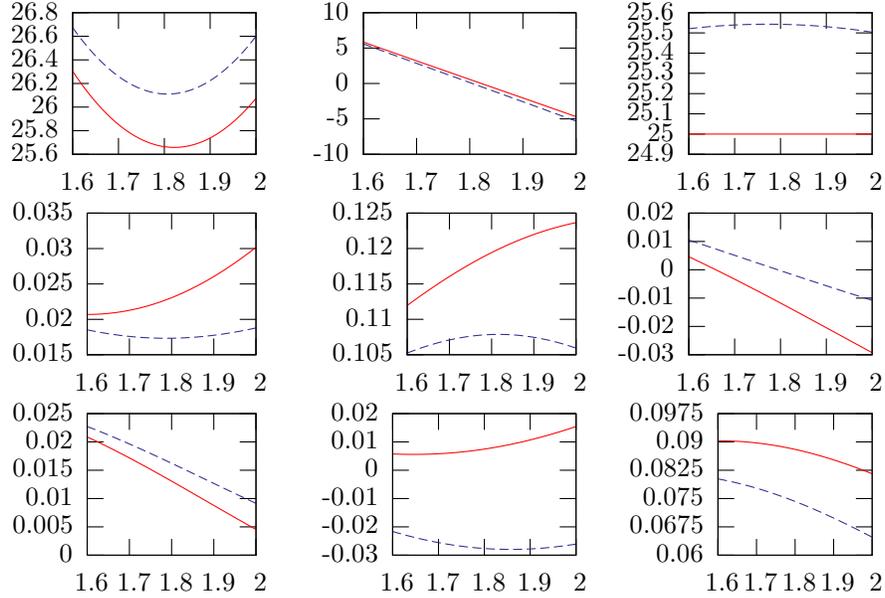}
  \caption{ \label{fig:2} The trajectories of all the variables over the bounce range, for the initial
conditions given in Eq.~(\ref{eq:ini_conditions}). The solid lines are the $W\left(\phi\right)=0$
case and the dashed lines correspond to the $W\left(\phi\right)=-0.05$ potential. From top left to
bottom right the trajectories are: $V\left(\phi\right)$, $\Ree J\left(\phi\right)$, $\Imm J\left(\phi\right)$,
$\Delta G^{VV}\left(\phi\right)$, $\Delta G^{J\bar{J}}\left(\phi\right)$,
$\Delta \Ree G^{VJ}\left( \phi\right)$, $ \Imm\Delta G^{VJ}\left(\phi\right)$, 
$\Delta \Ree G^{JJ}\left( \phi\right)$ and $ \Imm\Delta G^{JJ}\left(\phi\right)$. Notice in particular that
for $W\left(\phi\right)=0$ we have $\Imm J$ a constant, which is expected since for this case
we have $\Imm J = H_Q$ which 
is conserved and that several of the variables ($\Ree J\left(\phi\right)$, $\Delta G^{J\bar{J}}\left(
\phi\right)$ etc.) evolve asymmetrically around the bounce point.}
 \end{center}
\end{figure}

\begin{table}
\begin{center}
\begin{tabular}{|c|c|c|}
\hline\hline
$W\left(\phi\right)$ & $0$ & $-0.05$\\
\hline
$\frac{H_Q}{\rm vol} \Big|_{\rm Bounce}$ & 0.97438 & 1.01515\\
\hline
\end{tabular}
\caption{\label{tab:1} The kinetic energy density of the bounce depends on the magnitude and form of the
potential ($W\left(\phi\right)$) that is present. The initial conditions are given by Eq.~(\ref{eq:ini_conditions})}
\end{center}
\end{table}
For the case of $W\left(\phi\right) = -0.05$,
the trajectory $V\left(\phi\right)$ is plotted in
Fig.~\ref{fig:3}, with the initial conditions again given by Eq.~(\ref{eq:ini_conditions}). In this
case the kinetic energy density is found to be higher than the $W\left(\phi\right)=0$ case, see Table~\ref{tab:1}
(which gives the square-root of the energy densities).
Using the values of the moments at the bounce, we can evaluate the expected Hamiltonian density from
Eqs.~(\ref{eq:Ham_W1})--(\ref{eq:Ham_W1_approx}) and compare them to the numerical result, see Table~\ref{tab:2}.
\begin{table}
 \begin{center}
 \begin{tabular}{|c|c|c|c|}
  \hline\hline 
  Estimate &  Eq.~(\ref{eq:Ham_W1}), $\delta \theta^2 \approx 0.0005$ & numerical \\
\hline
 $\frac{H_Q}{\rm vol}\Big|_{\rm Bounce}$ & 1.0038 & 1.0052 \\ \hline
 \end{tabular}
  \caption{ \label{tab:2} The energy density of the bounce for the $W\left(\phi\right)=-0.05$ case,
  calculated using the
  approximate analytic results Eq.~(\ref{eq:Ham_W1}) and Eq.~(\ref{eq:Ham_W1_approx}) compared to the numerical
  result.}
 \end{center}
\end{table}
Notice that for this case the evolution equations given in Appendix~\ref{sec:app} are only valid up to
${\mathcal O}\left(W^2\right)=2.5\times 10^{-3}$ and an accuracy beyond $\sim 3$ significant figures would
require the use of higher orders in the expansion. We can see that the approximation made in order to
solve Eq.~(\ref{eq:bounce_ed_W}) is accurate to, approximately,
$0.5\%$. 
It is also worth noting that 
the moments, as seen in Figure~\ref{fig:2}, do not evolve symmetrically around the bounce point.
In particular $\Ree J$, $\Delta
G^{J\bar{J}}$, $\Delta \Ree G^{VJ}$, $ \Imm\Delta G^{VJ}$ and $ \Imm\Delta G^{JJ}$ all decrease monotonically
across the bounce (for this system). 
This may be significant for deparametrizing the system in order to use
one of these variables as an internal time parameter, as suggested in
\cite{Recollapse}.

\begin{figure}
 \begin{center}
  \includegraphics{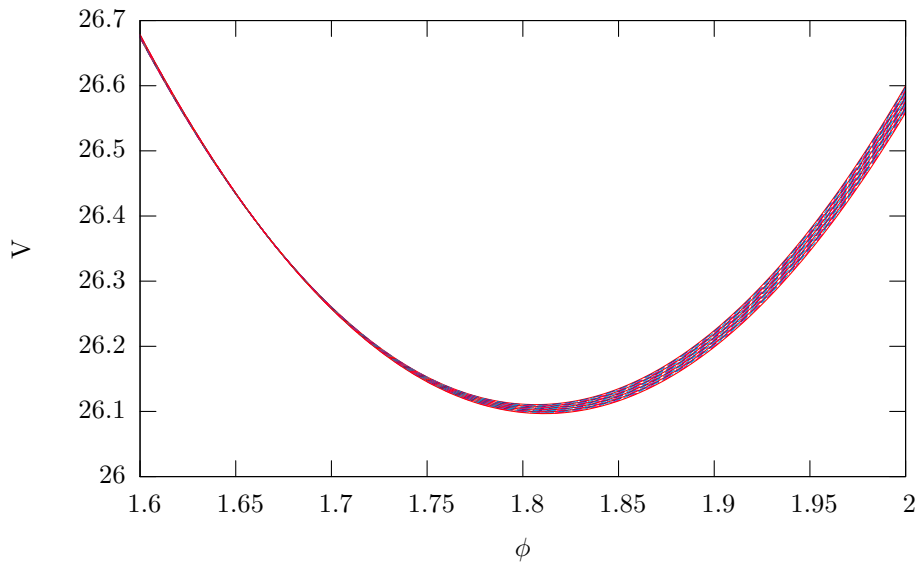}
  \caption{ \label{fig:4} The trajectory $V\left(\phi\right)$, evolved numerically for the initial conditions
given by Eq.~(\ref{eq:ini_conditions}). Again only the bounce region is plotted. In this case, the potential
was set to be $-0.05$. The various plots correspond to different initial values of $\Delta G_{J\bar{J}}$
in the range $\Delta G^{J\bar{J}}\Big|_{\rm bounce} = 0.107779\ \rightarrow\ 0.18926$
($G_{J\bar{J}}|_{\rm initial} = 100.0\ \rightarrow\ 140.0$, in steps
of $\delta G_{J\bar{J}} = 1.0$).  The sensitivity to variations of the
moments is significantly increased after the bounce, an observation
extending the conclusions of \cite{BeforeBB,Harmonic}.}
 \end{center}
\end{figure}

\begin{figure}
 \begin{center}
  \includegraphics{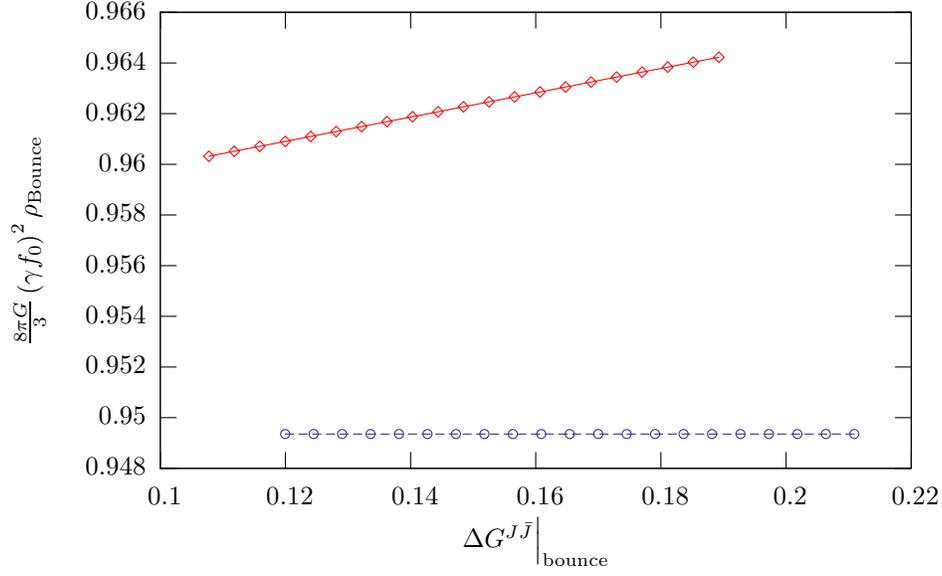}
  \caption{ \label{fig:5} The energy density of the bounce, given by Eq.~(\ref{eq:energy_den}),
 for the $W=0$ and $W=-0.05$  cases as a function of $\Delta G^{J\bar{J}}$, with the remaining
initial conditions being given by Eq.~(\ref{eq:ini_conditions}). The $W=0$ case (lower line) 
is approximately independent of $\Delta G^{J\bar{J}}$ because
at the bounce $\Delta G^{J\bar{J}} \sim \Delta G^{VV}$, for the system under consideration, however
for $W\neq 0$ (upper line) there is a (small) dependence on $G^{J\bar{J}}$. Note that both plots correspond 
to the same initial range of $G^{J\bar{J}} = 100 \rightarrow 140$.
The non-vanishing slope shows that effects on the bounce density are not just due to the non-vanishing potential.}
 \end{center}
\end{figure}

For the case of constant $W$ as studied here, the effect of varying
the moments is small but observable (for the evolutions considered
here). In Figure~\ref{fig:4} the trajectories $V\left(\phi \right)$
for various initial $\Delta G^{J\bar{J}}$ in the range, $\Delta
G^{J\bar{J}} \big|_{\rm bounce} = ( 0.108, 0.189)$. For
$W\left(\phi\right)=-0.05$ the effect of varying the initial moments
is even more pronounced than for $W=0$.  Figure~\ref{fig:5} shows the dependencies
of the (square-root of the) energy densities at the bounce on the
initial value of $\Delta G^{J\bar{J}}$, for both the
$W\left(\phi\right)=0$ and $W\left(\phi\right) =-0.05$ cases. In the
cases examined here, $\Delta G^{J\bar{J}}$ is approximately
proportional to $\Delta G^{VV}$ at the bounce. Hence, by
Eq.~(\ref{eq:W=0}), we would not expect there to be a significant
dependence on $\Delta G^{J\bar{J}}$ for the $W=0$ case, which is what
is found. For the $W\neq 0$ cases however there is a dependence, due
to the terms proportional to $W$ in Eq.~(\ref{eq:Ham_W1}).

In Section~\ref{sec:dyn_coherence} we described how to explicitly construct dynamical coherent states
for the $W\left(\phi\right)=0.0$ case. Given the initial data,
\be
 \left( V,\ \Imm J,\ G^{J\bar{J}},\ \Ree G^{JJ} \right) = \left( 105.0,\ 100.0,\ 125.0,\ -1.0\right)~,
\ee
we can solve Eq.~(\ref{eq:uncert_under}), with $U_1=U_2=U_3=0.0$ (which is required if the state is to be
dynamically coherent)  and Eq.~(\ref{eq:cond_4})  to find (for example)
\be
 \left( G_{VV},\ \Ree J,\ \Ree G_{VJ},\ \Imm G_{VJ},\ \Imm G_{JJ}\right) \approx
\left( 82.52970,\ 32.985295,\ 51.15507,\ 70.195184,\ 67.035438\right)~.
\ee
We have checked carefully that for this case (and others that are expected to be
dynamically coherent) the uncertainty relations are indeed conserved up to numerical
rounding errors (which can be made negligibly small by employing a suitably accurate discretization
scheme). This provides a strong test of the accuracy of our numerical implementation

\begin{figure}
 \begin{center}
  \includegraphics{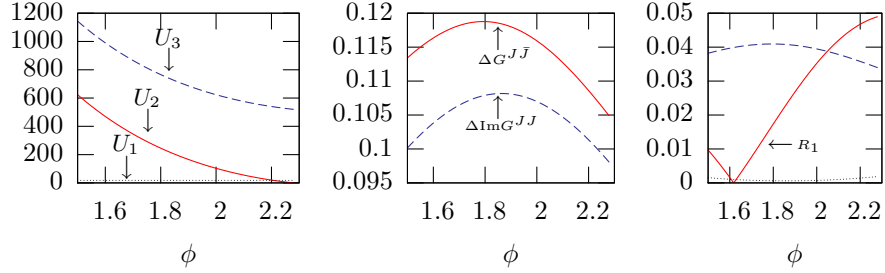}
  \caption{ \label{fig:uncert_breach} The left hand plot gives the values of the uncertainty conditions, 
during the evolution,
the center one plots the corresponding values of $\Delta G^{J\bar{J}}$ and $ \Imm\Delta G^{JJ}$.
The right hand plot shows $R_1$, the order of the terms neglected in
Eqs.~(\ref{eq:RealVV})--(\ref{eq:RealVJ})
compared to the fractional moments $(\Delta G^{J\bar{J}})^{3/2}$ (the
larger values) and $(\Delta G^{VV})^{3/2}$ (which happens to be much
smaller in this evolution).
For this evolution all the remaining moments and correlations
remain small (less than $\approx 2\%$) and the bounce occurs at $\phi_{\rm bounce} \approx 1.82$.
The increase of $R_1$ toward the end of the evolution indicates
that neglecting the third order moments is no longer a
good approximation in the region in which the uncertainty conditions (in this case $U_2$) are breached.}
 \end{center}
\end{figure}

Finally as an example of moments which evolve to break the uncertainty relations (in this particular
case Eq.~(\ref{eq:uncert_s_2})) showing the breakdown of the approximation, in Fig.~\ref{fig:uncert_breach} we plot the evolution of
$U_1$, $U_2$ and $U_3$ for the initial conditions
\be
\left( \begin{array}{ccccccccc}
 V & \Ree J & \Imm J & \Delta G^{VV} & \Delta G^{J\bar{J}} & \Delta \Ree G^{VJ} &  \Imm\Delta G^{VJ}
&  \Ree\Delta G^{JJ} &  \Imm\Delta G^{JJ} \\
35.0 & 24.0 & 25.0 & 0.0331 & 0.09159 & 0.04267 & 0.02363 & 0.0250 & 0.0737
\end{array}\right)~.
\ee
This initial state initially undersaturates all the uncertainty conditions ($U_1 \approx 18.1$, 
$U_2\approx 2623.9$ and $U_3\approx 2103.6$ at $\phi=1$), but dynamically evolves to
break the uncertainty bounds. As one can see from Fig.~\ref{fig:uncert_breach} the moments are
no longer small for this state ($\Delta G^{J\bar{J}}$ and $ \Imm\Delta
G^{JJ}$ are larger than $10\%$), which indicates that the contributions to the dynamics from higher order
moments can no longer be neglected. Note that even for this case, there is a bounce at 
$\phi \approx 1.82$, with square root of the energy density being $\rho_{\rm bounce}
\approx 0.97435$ and despite the fact that the evolution of this state has significant contribution from
higher order moments, this value is within $0.72\%$ of that predicted by Eq.~(\ref{eq:energy_bounce_W0}).
Thus we see that our estimates
for the (square root of the) energy density at the bounce
are well approximated using only the first order moment, at least while the dynamics of the 
state do not breach the uncertainty conditions Eq.~(\ref{eq:uncert_s_1})--(\ref{eq:uncert_s_3}).
Once again we emphasize that this breaching of the uncertainty conditions is not an artifact of the
numerical implementation, rather it is a genuine feature of the dynamics, due to the
truncation at second order in the moments.

%
%
%%%%%%%%%%%%%%%%%%%%%%%%%%%%%%%%%%%%%%%%%%%%%%%%%%%%%%%%%%%%%%%%%%%%%%%%%%%%%%%%%%%%%%%%%%%%%%%%%%%%%%%
%
%
\section{Conclusions}

Numerical investigations in loop quantum cosmology have so far only
considered semi-classical states of Gaussian type, peaked around some
particular value \cite{APS,APSCurved,NegCosNum}. The expectation value
of such states has been shown to closely follow the classical
trajectories, except at very high densities where quantum gravity
effects, but primarily those of quantum geometry, produce a
bounce. Implicit in these analyzes is the idea that a general
semi-classical state is well characterized by its expectation values
and small spreads alone and that the evolution of the spread and other
moments does not affect the evolution of the peak.  Here we have
considered a moment expansion of a state around such exactly solvable
Gaussian states as a way of probing the consequences of relaxing this
assumption. In principle one should consider all orders of the moment
expansion, however in order to make the system tractable, we have
truncated the series at second order.  Already this expands the
configuration space from $4$-dimensional ($(V,\phi)$ and their
momenta) to $18$-dimensional (given by the eight independent variables
in Eq.~(\ref{eq:variables}), $\phi$ and their momenta).

The inclusion of moments introduces sets of non-trivial uncertainty relations that must be satisfied
by the state {\it at all points along its trajectory}. Dynamical coherent states
are those for which all the uncertainty relations are exactly saturated, throughout the 
state's evolution. We have shown that the conditions necessary to ensure a state is dynamical
coherent, effectively reduces the dimension of the state's phase space i.e.\ the conditions
form a constraint surface on which the states, and their trajectories, lie. For the zero
potential case, there is only one condition that needs to be met to ensure that a state
which initially saturates the uncertainty bounds will be a dynamical coherent state
(i.e.\ will continue to saturate the bounds). For a system with a non-zero potential
one has four independent constraints (in addition to the uncertainty bounds) 
which need to be met, just to ensure that the first derivatives of the uncertainty
relations vanish. While requiring that the higher derivatives also vanish is likely
to introduce even more constraints. This makes it extremely difficult (and probably
impossible) to find exact dynamical coherent states for systems with a non-zero
scalar field potential. While conceptually this may be a problem, especially for considering
evolution from the infinite past ($\phi\rightarrow -\infty$) or 
cyclic models, in practice we are used to dealing with quantum states that are only
approximately dynamically coherent, such as particle wave-packets in
standard quantum mechanics.
Provided the states are approximately dynamically coherent for a sufficiently long
period of time ($\phi$) they can, for all practical purposes, be considered semi-classical.

We have investigated the consequences of the moments on the presence
of a bounce, for states that remain semi-classical (in the
sense that all their moments are small compared to the expectation
values and that the states obey the uncertainty conditions). In all
cases the bounce was found to be present and the energy density was
changed only marginally from the standard case (i.e.\ the case in
which the moments are neglected).  Thus we found that for the (large
number of) cases investigated, the bounce occurs at approximately the
same density, a feature not at all unexpected due to the setup of our
approximations. The fact that there were small deviations from the
standard bound is however significant, especially since we show that
it is possible for the bounce density to be sometimes larger
than the standard case. We restricted ourselves to considering only
states that remained highly semi-classical before, after and during
the bounce and still found that the moments noticeably back-react on
the trajectories of the expectation values. It may be that there are
states which are semi-classical at large scales, but which become
dominated by the evolution of their moments at small scales (i.e.\
become highly non-semi-classical) and for such states the presence of
a bounce is not guaranteed by current results in the literature.
%
%
%%%%%%%%%%%%%%%%%%%%%%%%%%%%%%%%%%%%%%%%%%%%%%%%%%%%%%%%%%%%%%%%%%%%%%%%%%%%%%%%%%%%%%%%%%%%%%%%%%%%%%%
%
%
\begin{acknowledgments}
 This work was supported in part by NSF grants PHY0748336,
PHY0854743, The George A.\ and Margaret M.~Downsbrough Endowment and
the Eberly research funds of Penn State. DM is supported by STFC, and
RT thanks the hospitality of Tokyo University of Science, University
of Tokyo and Rikkyo University, where this work was finalized during a
visit supported by a Royal Society bilateral grant.
\end{acknowledgments}

\begin{appendix}
\section{Evolution equations}
\label{sec:app}

The evolution of the variables $V$, $J$ and their moments
with respect to the scalar field $\phi$ are~\cite{BouncePot}:

\begin{eqnarray}
\frac{{\rm d}V}{{\rm d}\phi} &=& -\frac{J+\bar{J}}{2}+
\frac{J+\bar{J}}{(J-\bar{J})^2} V^{3/(1-x)} W(\phi)+ 3
\frac{J+\bar{J}}{(J-\bar{J})^4}V^{3/(1-x)}
(G^{JJ}+G^{\bar{J}\bar{J}}-2G^{J\bar{J}})W(\phi)
 \\
\nonumber && -\frac{6}{1-x} \frac{J+\bar{J}}{(J-\bar{J})^3}V^{(2+x)/(1-x)}
(G^{VJ}- G^{V\bar{J}})W(\phi)+
\frac{3}{2} \frac{2+x}{(1-x)^2} \frac{J+\bar{J}}{(J-\bar{J})^2}V^{(1+2x)/(1-x)}
G^{VV}\,W(\phi)
\\
&&- \frac{2V^{3/(1-x)}}{(J-\bar{J})^3} (G^{JJ}-G^{\bar{J}\bar{J}})W(\phi)+
\frac{3}{1-x}\frac{V^{(2+x)/(1-x)}}{(J-\bar{J})^2} 
(G^{VJ}+G^{V\bar{J}})W(\phi)\nonumber
\end{eqnarray}

\begin{eqnarray}
\frac{{\rm d}J}{{\rm d}\phi} &=& -(V+\hbar/2) + \left(\frac{3}{1-x}
\frac{V^{(2+x)/(1-x)}J}{J-\bar{J}}+
\frac{2V^{3/(1-x)}(V+\hbar/2)}{(J-\bar{J})^2}\right)
W(\phi)\label{eq-mot-loopJ}\\
&&+
\left(\frac{3}{1-x}\frac{V^{(2+x)/(1-x)}J}{(J-\bar{J})^3}+
\frac{6V^{3/(1-x)}(V+\hbar/2)}{(J-\bar{J})^4}\right)
(G^{JJ}+G^{\bar{J}\bar{J}}-2G^{J\bar{J}})W(\phi)\nonumber\\
&&-
3\left(\frac{2+x}{(1-x)^2}\frac{V^{(1+2x)/(1-x)}J}{(J-\bar{J})^2}+
\frac{4}{1-x}\frac{V^{(2+x)/(1-x)}(V+\hbar/2)}{(J-\bar{J})^3}\right)
(G^{VJ}-G^{V\bar{J}})W(\phi)\nonumber\\
&&+
\frac{3}{2}\left(\frac{(2+x)(1+2x)}{(1-x)^3}\frac{V^{3x/(1-x)}J}{J-\bar{J}}+
\frac{2(2+x)}{(1-x)^2}\frac{V^{(1+2x)/(1-x)}(V+\hbar/2)}{(J-\bar{J})^2}\right)
G^{VV}W(\phi)\nonumber\\
&&+ 4\frac{V^{3/(1-x)}}{(J-\bar{J})^3}(G^{V\bar{J}}-G^{VJ})W(\phi)
\nonumber\\
&&-
\frac{3}{1-x}\frac{V^{(2+x)/(1-x)}}{(J-\bar{J})^2} 
(G^{JJ}-G^{J\bar{J}}-2G^{VV})W(\phi)+
3\frac{2+x}{(1-x)^2}\frac{V^{(1+2x)/(1-x)}}{J-\bar{J}}G^{VJ}W(\phi) \nonumber
\end{eqnarray}

\begin{eqnarray} \label{EOMGpp}
\frac{{\rm d}G^{VV}}{{\rm d}\phi} &=& -(G^{VJ}+G^{V\bar{J}})+\frac{2V^{3/(1-x)}}{(J-\bar{J})^2}
(G^{VJ}+G^{V\bar{J}})W(\phi)\nonumber\\
&&- \frac{4V^{3/(1-x)}}{(J-\bar{J})^3}
\left(JG^{VJ}-\bar{J}G^{V\bar{J}}-
JG^{V\bar{J}}+\bar{J}G^{VJ}\right)W(\phi)\nonumber\\
&&+\frac{6}{1-x}\frac{V^{(2+x)/(1-x)}(J+\bar{J})}{(J-\bar{J})^2}G^{VV}W(\phi)-
\frac{\hbar^2}{2(1-x)}\frac{V^{(2+x)/(1-x)}(J+\bar{J})}{(J-\bar{J})^2}W(\phi)\nonumber\\
&=&-\left(1/W(\phi)-\frac{2V^{3/(1-x)}}{(J-\bar{J})^2}\right)
(G^{VJ}+G^{V\bar{J}})W(\phi)- \frac{4V^{3/(1-x)}(J+\bar{J})}{(J-\bar{J})^3}
(G^{VJ}-G^{V\bar{J}})W(\phi)\nonumber\\
&&+\frac{6}{1-x}\frac{V^{(2+x)/(1-x)}(J+\bar{J})}{(J-\bar{J})^2}G^{VV}W(\phi)-
\frac{\hbar^2}{2(1-x)}\frac{V^{(2+x)/(1-x)}(J+\bar{J})}{(J-\bar{J})^2}W(\phi)
\end{eqnarray}

\begin{eqnarray} \label{EOMGJbarJ}
\frac{{\rm d}G^{J\bar{J}}}{{\rm d}\phi} &=& -(G^{VJ}+G^{V\bar{J}})+
\frac{2V^{3/(1-x)}}{(J-\bar{J})^2}
(G^{VJ}+G^{V\bar{J}})W(\phi)- \frac{4V^{3/(1-x)}(V+\hbar/2)}{(J-\bar{J})^3}
(G^{JJ}-G^{\bar{J}\bar{J}})W(\phi)\nonumber\\
 &&- \frac{3}{1-x}\frac{V^{(2+x)/(1-x)}}{(J-\bar{J})^2}
((J+\bar{J})G^{J\bar{J}}- (2V+\hbar)(G^{VJ}+G^{V\bar{J}})-
\bar{J}G^{JJ}-JG^{\bar{J}\bar{J}}\nonumber\\
&&+{\textstyle\frac{1}{6}}\hbar^2(J+\bar{J})) W(\phi)
+\frac{3(2+x)}{(1-x)^2}\frac{V^{(1+2x)/(1-x)}}{J-\bar{J}}(JG^{V\bar{J}}-
\bar{J}G^{VJ})W(\phi)\nonumber\\
&=& -(G^{VJ}+G^{V\bar{J}})\nonumber\\
&&+\left(\frac{3}{2}\frac{2+x}{(1-x)^2}V^{(1+2x)/(1-x)}+
\frac{2V^{(2+x)/(1-x)}(V+3(V+\hbar/2)/(1-x))}{(J-\bar{J})^2}\right)
 (G^{VJ}+G^{V\bar{J}})W(\phi)\nonumber\\
&& -\frac{3}{2}\frac{2+x}{(1-x)^2}\frac{V^{(1+2x)/(1-x)}
(J+\bar{J})}{J-\bar{J}} 
(G^{VJ}-G^{V\bar{J}})W(\phi)\nonumber\\
&& - \frac{3}{2} \frac{V^{(2+x)/(1-x)}}{J-\bar{J}}
\left(\frac{1}{1-x}+\frac{8}{3}\frac{V(V+\hbar/2)}{(J-\bar{J})^2}\right) 
(G^{JJ}-G^{\bar{J}\bar{J}})W(\phi)\nonumber\\
&& +\frac{3}{2(1-x)} 
\frac{V^{(2+x)/(1-x)}(J+\bar{J})}{(J-\bar{J})^2}
(G^{JJ}+G^{\bar{J}\bar{J}}- 2G^{J\bar{J}})W(\phi)
-\frac{\hbar^2}{2(1-x)} \frac{V^{(2+x)/(1-x)}(J+\bar{J})}{(J-\bar{J})^2}W(\phi)
\end{eqnarray}

\begin{eqnarray} \label{EOMGpJ}
\frac{{\rm d}G^{VJ}}{{\rm d}\phi} &=& -\frac{1}{2}(G^{JJ}+G^{J\bar{J}}+2G^{VV})+
\frac{V^{3/(1-x)}}{(J-\bar{J})^2}
(G^{JJ}+G^{J\bar{J}}+2G^{VV})W(\phi)
+\frac{3}{1-x}\frac{V^{(2+x)/(1-x)}}{J-\bar{J}}G^{VJ}W(\phi)\\
&&+\frac{V^{3/(1-x)}W(\phi)}{(J-\bar{J})^3} 
(-2(J+\bar{J})G^{JJ}-2JG^{\bar{J}\bar{J}}
+2(V+\hbar/2)(4G^{V\bar{J}}-2G^{VJ}) +2JG^{J\bar{J}}\nonumber\\
&&
-\hbar^2\bar{J}+{\textstyle \frac{1}{3}}\hbar^2J)
+\frac{3}{1-x}\frac{V^{(2+x)/(1-x)}W(\phi)}{(J-\bar{J})^2}
(6(V+\hbar/2)G^{VV}-\bar{J}G^{VJ}-JG^{V\bar{J}}
+{\textstyle\frac{1}{2}}\hbar^2V+{\textstyle\frac{1}{4}}\hbar^3)\nonumber\\
&&+
3\frac{2+x}{(1-x)^2}\frac{V^{(1+2x)/(1-x)}W(\phi)}{J-\bar{J}} 
(JG^{VV}-{\textstyle\frac{1}{12}}\hbar^2J)\nonumber\\
&=& \overline{\dot{G}^{V\bar{J}}} \nonumber
\end{eqnarray}

\begin{eqnarray}\label{EOMGJJ}
\frac{{\rm d}G^{JJ}}{{\rm d}\phi} &=& -2G^{VJ}+
4\frac{V^{3/(1-x)}}{(J-\bar{J})^2}G^{VJ}W(\phi)
+\frac{6}{1-x}\frac{V^{(2+x)/(1-x)}}{J-\bar{J}}G^{JJ}W(\phi)\\
&&+ \frac{V^{3/(1-x)}W(\phi)}{(J-\bar{J})^3}
(8(V+\hbar/2)(2G^{VV}+ G^{J\bar{J}}-G^{JJ})- 8JG^{V\bar{J}}
-8\bar{J}G^{VJ} +4\hbar^2V+2\hbar^3) \nonumber\\
&&-\frac{3}{1-x}\frac{V^{(2+x)/(1-x)}W(\phi)}{(J-\bar{J})^2} 
(2JG^{JJ}- 8(V+\hbar/2)G^{VJ}
+2\bar{J}G^{JJ}+\hbar^2J)
+ \frac{6(2+x)}{(1-x)^2} \frac{V^{(1+2x)/(1-x)}}{J-\bar{J}} \nonumber \\
%\frac{12V}{J-\bar{J}}JG^{VJ}W(\phi)\nonumber\\
&=& \overline{\dot{G}^{\bar{J}\bar{J}}}\,. \nonumber
\end{eqnarray}

Equations of motion for real and imaginary parts of $J$, $G^{VJ}$ and
$G^{JJ}$ are:
\begin{eqnarray}
 \frac{\md}{\md\phi} {\rm Re}J &=& -\left( V+\frac{\hbar}{2}\right) +\frac{1}{2}\left(\frac{3}{1-x}
V^{(2+x)/(1-x)}-
\frac{V^{(4-x)/(1-x)}}{({\rm Im}J)^2}\right)
W(\phi)\\
&&+
\frac{3}{4}\frac{V^{(4-x)/(1-x)}}{({\rm Im}J)^4}
({\rm Re}G^{JJ}-G^{J\bar{J}})W(\phi)\nonumber\\
&&+
\left(\frac{3}{(1-x)}-1\right)
\frac{V^{3/(1-x)}}{({\rm Im}J)^3}
{\rm Im}G^{VJ}W(\phi)\nonumber\\
&&+
\frac{3}{2}\left(\frac{(2+x)(1+2x)}{2(1-x)^3}V^{3x/(1-x)}-
\frac{4-x}{2(1-x)^2}\frac{V^{(2+x)/(1-x)}}{({\rm Im}J)^2}\right)
G^{VV}W(\phi)\nonumber \\
&& + \frac{\hbar}{2}\frac{ W(\phi)}{ \left(\Imm J\right)^2 } V^{3/(1-x)}
\Bigl[ \frac{-1}{2} +\frac{3}{4\left(\Imm J\right)^2} \left( \Ree G^{JJ} - G^{J\bar{J}}\right)
\nonumber \\
&&+\frac{3}{(1-x)\Imm J}\Imm G^{VJ} V^{-1} - \frac{3(2+x)}{4(1-x)^2} G^{VV} V^{-2} \Bigr]~,
\end{eqnarray}

\begin{eqnarray}
 \frac{\md}{\md\phi}{\rm Im}J &=& -\frac{3}{2(1-x)}
\frac{V^{(2+x)/(1-x)}{\rm Re}J}{{\rm Im}J} W(\phi)-
\frac{3}{4}\frac{(2+x)(1+2x)}{(1-x)^3}\frac{V^{3x/(1-x)}{\rm Re}J}{{\rm Im}J}
G^{VV}W(\phi)\\
&&+
\frac{3}{4(1-x)}\frac{V^{(2+x)/(1-x)}{\rm Re}J}{({\rm Im}J)^3}
({\rm Re}G^{JJ}-G^{J\bar{J}})W(\phi)+
\frac{3}{4(1-x)}\frac{V^{(2+x)/(1-x)}}{({\rm Im}J)^2} 
{\rm Im}G^{JJ}W(\phi)\nonumber\\
&&+
\frac{3}{2}\frac{2+x}{(1-x)^2}\frac{V^{(1+2x)/(1-x)}{\rm Re}J}{({\rm Im}J)^2}
{\rm Im}G^{VJ}W(\phi)-
\frac{3}{2}\frac{2+x}{(1-x)^2}\frac{V^{(1+2x)/(1-x)}}{{\rm Im}J}
{\rm Re}G^{VJ}W(\phi)\nonumber
\end{eqnarray}

\begin{eqnarray}
 \frac{\md}{\md\phi}{\rm Re}G^{VJ} &=& -\frac{1}{2} ({\rm Re}G^{JJ}
 + G^{J\bar{J}}+ 2G^{VV}) \\
&& -\frac{1}{2}\frac{V^{3/(1-x)}}{({\rm Im}J)^2}
 \left(G^{J\bar{J}}+\frac{10-x}{1-x} G^{VV}\right)
 W(\phi)+ \frac{3}{2}\frac{2+x}{(1-x)^2}
 V^{(1+2x)/(1-x)} G^{VV} W(\phi)\nonumber\\
 && + \left(\frac{3}{2(1-x)}
   \frac{V^{(2+x)/(1-x)}}{{\rm Im}J}+ \frac{3}{2}
   \frac{V^{(4-x)/(1-x)}}{({\rm Im}J)^3}\right) {\rm Im}G^{VJ}
 W(\phi)\nonumber\\
&&+ \frac{3}{2(1-x)} \frac{V^{(2+x)/(1-x)}
   {\rm Re}J}{({\rm Im}J)^2}
   {\rm Re}G^{VJ} W(\phi)
 +\frac{1}{4}\frac{V^{3/(1-x)}{\rm Re}J}{({\rm Im}J)^3}
 {\rm Im}G^{JJ} W(\phi)\nonumber \\
&& + \frac{\hbar}{2} \frac{W(\phi) V^{3/(1-x)}}{\Imm J} \Biggl[ \frac{3}{2} \frac{\Imm G^{VJ}}{\left( \Imm J\right)^2}
-\frac{\hbar}{2} \left( \frac{2}{3\Imm J} + \frac{ 2+x}{2\left( 1-x\right)^2} V^{-2}\Ree J \right) \nonumber \\
&&- \frac{3}{1-x} \frac{V^{-1}}{ \Imm J} \Biggl( G^{VV} + \frac{\hbar}{4}\left( V+\frac{\hbar}{2}\right) \Biggr) \Biggr]~,
\end{eqnarray}

\begin{eqnarray}
 \frac{\md}{\md\phi}{\rm Im}G^{VJ} &=& -\frac{1}{2} {\rm Im} G^{JJ} -\frac{1}{2}
 \frac{V^{(2+x)/(1-x)}}{{\rm Im}J}\left(\frac{3}{1-x}
- \frac{V^2}{({\rm Im}J)^2}\right) {\rm Re}G^{VJ} W(\phi)\\
 && -\frac{3}{2} \frac{2+x}{(1-x)^2} \frac{V^{(1+2x)/(1-x)}{\rm
 Re}J}{{\rm Im}J} G^{VV}W(\phi)+ \frac{1}{4}\frac{V^{3/(1-x)}{\rm
 Re}J}{({\rm Im}J)^3} G^{J\bar{J}}W(\phi)\nonumber\\
 && -\frac{1}{4}\frac{V^{3/(1-x)}}{({\rm Im}J)^3} (3{\rm Re}J{\rm
 Re}G^{JJ} + {\rm Im}J{\rm Im}G^{JJ})W(\phi)\nonumber \\
&&+ \frac{\hbar}{2} W\left(\phi\right) V^{3/(1-x)}\Biggl[
\frac{\Ree G^{VJ}}{2\left(\Imm J\right)^3} -\frac{\hbar}{2} \left( \frac{\Ree J}{3\left(\Imm J\right)^3}
- \frac{2+x}{16\left( 1-x\right)^2} \frac{V^{-2}\Ree J}{\Imm J}\right) \Biggr]
\end{eqnarray}

\begin{eqnarray}
 \frac{\md}{\md\phi} {\rm Re} G^{JJ} &=&
 -2{\rm Re}G^{VJ}+\left(3\frac{2+x}{(1-x)^2} V^{(1+2x)/(1-x)}+
 \frac{7-x}{1-x} \frac{V^{3/(1-x)}}{({\rm Im}J)^2}\right) {\rm
 Re}G^{VJ}W(\phi)\\
&&+3\frac{2+x}{(1-x)^2} \frac{V^{(1+2x)/(1-x)}{\rm Re}J}{{\rm Im}J}
 {\rm Im}G^{VJ}W(\phi)\nonumber\\
&&+\frac{3}{1-x} \frac{V^{(2+x)/(1-x)} {\rm Re}J} {{\rm Im}J^2} 
 {\rm Re}G^{JJ}W(\phi) \nonumber\\
&& + \frac{V^{(4-x)/(1-x)}}{({\rm Im}J)^3} 
\left ( 1+ \frac{3}{(1-x)} \frac{({\rm Im}J^2)}{V^2} \right ) {\rm Im}G^{JJ}W(\phi)\nonumber \\
&& + \frac{\hbar}{2} \frac{ W\left(\phi\right) V^{3/(1-x)}}{\left(\Imm J\right)^2}
\left[ \frac{ \Imm G^{JJ}}{\Imm J} + \frac{12}{1-x} V^{-1} \left( 2\Ree G^{VJ}
-\frac{\hbar}{2} \Ree J \right)\right]~,
\end{eqnarray}

\begin{eqnarray}
 \frac{\md}{\md\phi} {\rm Im} G^{JJ} &=& -2{\rm Im}G^{VJ}
 -\left(\frac{5+x}{1-x}
   \frac{V^{3/(1-x)}}{({\rm Im}J)^2}+ \frac{6(2+x)}{2(1-x)^2}
   V^{(1+2x)/(1-x)}\right) {\rm Im}G^{VJ}
 W(\phi)\nonumber\\
&& -\left(3\frac{2+x}{(1-x)^2}+2
  \frac{V}{({\rm Im}J)^2} \right)
\frac{V^{(1+2x)/(1-x)} {\rm Re}J}{{\rm Im}J}
 {\rm Re}G^{VJ} W(\phi)\nonumber\\
&&+\frac{V^{(4-x)/(1-x)}}{({\rm Im}J)^3} (2G^{VV}+G^{J\bar{J}})
  W(\phi)  -\left(\frac{3}{1-x}+
  \frac{V^2}{({\rm Im}J)^2}\right)
\frac{V^{(2+x)/(1-x)}}{{\rm Im}J}
 {\rm Re}G^{JJ} W(\phi)\nonumber\\
&&+\frac{3}{1-x}
\frac{V^{(2+x)/(1-x)}{\rm Re}J}{({\rm Im}J)^2}
 {\rm Im}G^{JJ} W(\phi) \nonumber \\
&& -\frac{\hbar}{2} \frac{W\left(\phi\right)V^{3/(1-x)}}{\left(\Imm J\right)^3} \Biggl[
 2G^{VV} + G^{J\bar{J}} - \Ree G^{JJ} + 16\left(\frac{\hbar}{2}\right)^2 \left( V+\frac{\hbar}{2}\right)
\nonumber \\
&&-\frac{12 \Imm J V^{-1} }{1-x} \left( 2\Imm G^{VJ} - \frac{\hbar}{2} \Imm J \right)\Biggr]~.
\end{eqnarray}

%
%
%%%%%%%%%%%%%%%%%%%%%%%%%%%%%%%%%%%%%%%%%%%%%%%%%%%%%%%%%%%%%%%%%%%%%%%%%%%%%%%%%%%%%%%%%%%%%%%%%%%%%%%
%
%
\section{Fiducial volume}\label{sec:fi_vol}

For the mathematical formulation of cosmological models with
non-compact spatial slices, a fiducial volume $V_0$ is required in
order for the symplectic structure of the form $\int{\rm d}^3x
p_{\phi}\dot{\phi}$ to be defined (i.e.\ for spatial integrals to be
finite). Physical quantities must be independent of the value of
$V_0$ and in particular remain well defined as $V_0\rightarrow
\infty$.  This has, in the past, been the cause of some confusion, so
for clarity in this section we describe in detail the $V_0$ dependence
of the all the parameters in the theory.

The canonical pair we are using is,
\beq\label{eq:poisson}
 \left\{ f\left(p\right)\ c,V\right\} &=& \left\{ f_0 p^x c, \frac{3p^{1-x}}{8\pi G \gamma \left( 1-x\right)f_0}\right\}
\nonumber \\
&=& \frac{3}{8\pi \gamma G} \left\{ p,c\right\}
= \frac{3}{8\pi \gamma G} V_0 \left\{ \tilde{p},\tilde{c}\right\}=1~,
\eeq
where the over-tilde implies that the quantity {\it does not} depend on $V_0$.
The combination $f_0 p^x c$ thus behaves as
\be
 f_0 p^x c = \tilde{f}_0 \tilde{p}^x \tilde{c} V_0^{y + \frac{2x+1}{3}}~,
\ee
where $f_0 = \tilde{f}_0 V_0^y$. The freedom we have then is to fix the
behavior of $f_0$ with respect to $V_0$ such that a well-defined
formulation results. We could, for example, have chosen to
set $y = 0$, as implicitly occurred in some papers on the subject.
Had we done so, we would have found:
\be
 V = \frac{3 p^{1-x}}{8\pi G \gamma\left(1-x\right)f_0}
 = \frac{3 \tilde{p}^{1-x}}{8\pi G \gamma\left(1-x\right)\tilde{f}_0} 
V_0^{2(1-x)/3}\,.
\ee
However, then we  have
\be
 f(p)c=f_0 p^x c = \tilde{f}_0 \tilde{p}^x \tilde{c} V_0^{(1+2x)/3}~,
\ee
which only for $x=-1/2$ is consistent with the primary condition on the scaling
behavior of $f(p)$, ensuring that holonomies are independent of
$V_0$. In particular, if we were to attempt a limit of large $V_0$,
holonomies or even
observables such as the density at the bounce would not remain well-defined.

Instead, we must choose to have $f_0$ depend on $V_0$ as,
\be
f_0 = \tilde{f}_0 V_0 ^{-\frac{1+2x}{3}},
\ee
so that $f_0 p^x c = \tilde{f}_0 \tilde{p}^x\tilde{c}$ is independent
of the regulator $V_0$. The
parameter $f_0$ is $V_0$-independent only for $x=-1/2$, in which case
$f_0p^xc= f_0\gamma\dot{a}/a$ appears in holonomies, as suggested as a
special ad-hoc choice by~\cite{APS}. Then we have that
\be
V = \frac{ 3 \tilde{p}^{1-x}}{8\pi G \gamma \left( 1- x\right) \tilde{f}_0} V_0~.
\ee
The factor $V_0$ provides the correct dependence on $V_0$ in the
Poisson bracket, Eq.~(\ref{eq:poisson}).  In this case, taking the
large-$V$ limit corresponds either to taking the large-$V_0$
limit or the small-discreteness
limit ($\tilde{f}_0 \rightarrow 0$).

We are interested in physical quantities, such as the kinetic energy density at the bounce (here
we focus on the $W=0$ case, the $W \neq 0$ case follows similarly), the square-root of which
is given by Eq.~(\ref{eq:energy_bounce_W0}),
\beq
\frac{H_Q}{\rm vol} \Big|_{\rm Bounce,\ {\rm large}V}&=&
\left[ \frac{ 3\tilde{p}^{ -\left( 1+2x\right)/2} }{8\pi\gamma G\left( 1-x\right)} \tilde{f}_0
 \sqrt{\frac{ 1+\Delta G^{VV}}{1+\Delta G^{J\bar{J}}}}
 \tilde{p}^{-3/2}\right]_{\rm Bounce,\ {\rm large}V}~,
\eeq
which is independent of the fiducial volume for all choices of
the lattice refinement parameters $x$ and $\tilde{f}_0$.

\end{appendix}

%\bibliographystyle{prsty}
%\bibliography{../Bib/QuantGra}

\end{document}